\documentclass{emulateapj}
\usepackage{apjfonts}
\usepackage{amsmath}
\usepackage{multirow}

\slugcomment{Draft Version 25 July, 2017}

\shorttitle{Circumnuclear Disks in ETGs with ALMA}
\shortauthors{Boizelle et al.}

\usepackage{natbib}

\begin{document}

\title{ALMA Observations of Circumnuclear Disks in Early Type Galaxies: {}$^{12}$CO(2$-$1) and Continuum Properties}

\author{Benjamin D. Boizelle}
\affil{Department of Physics and Astronomy, 4129 Frederick Reines Hall, University of California, Irvine, CA, 92697-4575, USA; bboizell@uci.edu}

\author{Aaron J. Barth}
\affil{Department of Physics and Astronomy, 4129 Frederick Reines Hall, University of California, Irvine, CA, 92697-4575, USA}

\author{Jeremy Darling}
\affil{Center for Astrophysics and Space Astronomy, Department of Astrophysical
and Planetary Sciences, University of Colorado, 389 UCB, Boulder,
CO 80309-0389, USA}

\author{Andrew J. Baker}
\affil{Department of Physics and Astronomy, Rutgers, the State University
of New Jersey, 136 Frelinghuysen Road Piscataway, NJ 08854-8019, USA}

\author{David A. Buote}
\affil{Department of Physics and Astronomy, 4129 Frederick Reines Hall, University of California, Irvine, CA, 92697-4575, USA}

\author{Luis C. Ho}
\affil{Kavli Institute for Astronomy and Astrophysics, Peking University,
Beijing 100871, China; Department of Astronomy, School of Physics,
Peking University, Beijing 100871, China}

\author{Jonelle L. Walsh}
\affil{George P. and Cynthia Woods Mitchell Institute for Fundamental Physics and Astronomy, 4242 TAMU, Texas A\&M University, College Station, TX, 77843-4242, USA}

\begin{abstract}

We present results from an Atacama Large Millimeter/submillimeter Array (ALMA) \mbox{Cycle 2} program to map CO(2$-$1) emission in nearby early-type galaxies (ETGs) that host circumnuclear gas disks. We obtained $\sim0\farcs3-$resolution Band 6 observations of seven ETGs selected on the basis of dust disks in {\it Hubble Space Telescope} images. We detect CO emission in five at high signal-to-noise ratio with the remaining two only faintly detected. All CO emission is coincident with the dust and is in dynamically cold rotation. Four ETGs show evidence of rapid central rotation; these are prime candidates for higher-resolution ALMA observations to measure the black hole masses. In this paper we focus on the molecular gas and continuum properties. Total gas masses and H$_2$ column densities for our five CO-bright galaxies are on average $\sim10^8$ $M_\sun$ and $\sim10^{22.5}$ cm$^{-2}$ over the $\sim{\rm kpc}$-scale disks, and analysis suggests that these disks are stabilized against gravitational fragmentation. The continuum emission of all seven galaxies is dominated by a central, unresolved source, and in five we also detect a spatially extended component. The $\sim$230 GHz nuclear continua are modeled as power laws ranging from $S_\nu \sim \nu^{-0.4}$ to $\nu^{1.6}$ within the observed frequency band. The extended continuum profiles of the two radio-bright (and CO-faint) galaxies are roughly aligned with their radio jet and suggests resolved synchrotron jets. The extended continua of the CO-bright disks are coincident with optically thick dust absorption and have spectral slopes that are consistent with thermal dust emission.

\end{abstract}


\section{Introduction}
\label{intro}

An early consensus held that early-type galaxies (ETGs, encompassing elliptical and S0 galaxies) were nearly devoid of gas and dust \citep{hub26,dev59,san61}. The first challenge to this paradigm came from optical spectroscopy that revealed the presence of ionized interstellar gas (e.g., \citealp{may58,min59,ost60}; for a more recent survey, refer to \citealp{bas17}). Later, radio observations of ETGs detected neutral hydrogen on large (several kpc) scales (e.g., \citealp{bal72,kna85,war86}), and X-ray observations uncovered reservoirs of hot, diffuse gas within and around some cluster and isolated ETGs (e.g., \citealp{for79,nul84}). Soon afterwards, the {\it Infrared Astronomical Satellite} ({\it IRAS}) discovered significant mid to far-IR excess in ETGs originating from thermal dust emission (e.g., \citealp{jur86,kna89}). Imaging surveys from the ground and with the {\it Hubble Space Telescope} ({\it HST}) also detected dust in absorption in the centers of roughly half of all early-type galaxies (\citealp{ebn85,ebn88,van95, ver99, tom00, tra01, lai03,lau05}). In about 10\% of these ETGs, {\it HST} imaging revealed round, morphologically regular disks (typically with sub-kpc radii) that trace a dense, cold component of the nuclear environment. Early CO observations of several early-type targets with bright far-IR emission confirmed the presence of molecular gas with $\sim50\%$ detection rates (e.g., \citealp{sag89, wik95}). While at much coarser spatial resolution than {\it HST} imaging, these gas observations suggested that up to a tenth of ETGs might possess molecular gas that is both detectable in CO emission and in disk-like rotation on small scales.

Two prominent surveys -- SAURON \citep{dez02} and ATLAS$^{\rm 3D}$ \citep{cap11} -- observed CO in representative \citep{com07} and volume-limited \citep{you11,ala13} samples of nearby ETGs, respectively. In the latter survey, \citet{you11} detect CO emission in about a quarter of their galaxies, and \citet{you14} determine that $\sim40\%$ of nearby ETGs harbor significant ($\gtrsim10^8$ $M_\odot$) molecular and/or atomic gas reservoirs. Nearly half of the ATLAS$^{\rm 3D}$ galaxies with interferometric detections (just over 10\% of their full sample) have disk-like CO morphologies, while the rest show ring-like and non-axisymmetric CO structures \citep{ala13}. Even at moderate resolution (beam FWHM $\gtrsim3\arcsec$, corresponding to $\sim{\rm kpc}$ scales), these mm-wavelength interferometric observations demonstrate that disk-like CO morphologies tend to correlate with regular gas rotation in ETGs. The gas kinematic axes are typically aligned with the galaxy photometric axes, although often with moderate ($\sim30\degr$) disk warping. Higher resolution ($\sim0\farcs25$) observations confirm low turbulent velocity dispersions ($\sim10$ km s$^{-1}$) on scales of a few tens of parsecs from the galaxy centers (e.g., \citealp{uto15}). ATLAS$^{\rm 3D}$ galaxies with regular, circularly rotating molecular gas disks tend to have coincident, morphologically round dust features, indicating that dust morphology is tied to the dynamical coldness of the underlying gas disk.

Nearly two decades of supermassive black hole (BH) mass measurements have revealed correlations between the central BH mass and large-scale host galaxy properties (e.g., the $M_{\rm BH}-\sigma_\star$ relationship; \citealp{geb00,fer00}). A better understanding of BH demographics, based on a growing sample of BH masses, has shown a more complicated connection between these central masses and properties of their host galaxy, such as classical bulges vs.\ pseudobulges and cored vs.\ coreless ellipticals \citep{kor13}. These BH masses are typically measured by dynamically modeling resolved stellar or gaseous kinematics. Both methods are subject to potentially large systematic uncertainties (e.g., \citealp{van10,wal10}), although gas dynamical modeling is conceptually and computationally simpler than modeling the stellar orbital structure of an entire galaxy. Dynamically cold, dusty disks in ETGs are therefore appealing targets for BH mass measurements. Disentangling the kinematic signature of the BH from the host galaxy requires resolving the dynamical sphere of influence $r_{\rm g}\approx {\rm G} M_{\rm BH}/\sigma_\star^2$. Within this radius, the BH mass dominates over the extended stellar mass and gives rise to an elevated central stellar velocity dispersion $\sigma_\star$. For luminous ETGs, $r_{\rm g}$ is usually on the order of a few tens of parsecs; at a distance of $\sim20$ Mpc, this corresponds to up to a few tenths of an arcsecond.

\begin{deluxetable*}{lccccccc}[h!]
\tabletypesize{\scriptsize}
\tablecaption{Galaxy Sample Characteristics\label{tbl-obspars}}
\tablewidth{0pt}
\tablehead{
\multirow{2}{*}{\phantom{m}Galaxy} & \multirow{2}{*}{Type} & \colhead{$M_{K_s}$} & \colhead{$\sigma_\star$} & \colhead{$D_L$} & \colhead{$v_{\rm sys}$} & \colhead{Angular Scale} & \colhead{$b$ / $a$ (dust)} \\
 & & \colhead{(mag)} & \colhead{(km s$^{-1}$)} & \colhead{(Mpc)} & \colhead{(km s$^{-1}$)} & \colhead{(pc arcsec$^{-1}$)} & \colhead{($''$)} \\
\phantom{mag}(1) & (2) & \multicolumn{1}{c}{(3)} & (4) & \multicolumn{1}{c}{(5)} & (6) & \multicolumn{1}{c}{(7)} & \multicolumn{1}{c}{(8)}
 }
\startdata
NGC 1332 & S0 & $-24.69$ (0.17) & 313 & 22.3 (1.8) & 1524 & 106.9 & 0.28 / 2.1 \\
\rule{0pt}{4ex}NGC 1380 & SA0 & $-24.30$ (0.17) & 211 & 17.1 (1.4) & 1877 & \phantom{ }82.1 & \phantom{ }1.5 / 4.8 \\
\rule{0pt}{4ex}NGC 3258 & E1 & $-24.21$ (0.25) & 257 & 31.9 (3.9) & 2792 & 148.4 & \phantom{ }0.7 / 1.1 \\
\rule{0pt}{4ex}NGC 3268 & E2 & $-24.50$ (0.24) & 235 & 33.9 (3.9) & 2800 & 161.2 & \phantom{ }1.2 / 2.4 \\
\rule{0pt}{4ex}NGC 4374 (M84) & E1 & $-25.04$ (0.11) & 275 & 17.9 (0.9) & 1017 & \phantom{ }86.0 & 0.35 / 1.5 \\
\rule{0pt}{4ex}NGC 6861 & SA0 & $-24.47$ (0.33) & 414 & 27.3 (4.5) & 2829 & 129.8 & 1.95 / 8.0 \\
\rule{0pt}{4ex}IC 4296 & E & $-25.88$ (0.16) & 329 & 47.4 (3.7) & 3737 & 213.2 & 0.25 / 0.8 \\
\enddata
\tablecomments{Col. (2): Galaxy type, taken from \citet{dev91}. Col. (3): $K_s-\,$band magnitude and uncertainty from the 2MASS catalogue \citep{skr06}, including the uncertainty in the luminosity distance $D_L$. Col. (4): Central stellar velocity dispersion from the HyperLeda database \citep{mak14}. Col. (5): Luminosity distance and uncertainties as measured from surface brightness fluctuations by \citet{mei00} and \citet{ton01}, including a Cepheid zero-point correction \citep{mei05}. Uncertainties do not incorporate any systematic uncertainties in this zero-point. Col. (6): Systemic velocity $v_{\rm sys}$ taken from the NASA/IPAC Extragalactic Database (NED). Col. (7): Angular scale derived from the assumed $D_L$. Col. (8): Dust disk minor/major axis radii measured from two-band {\it HST} color images. For \mbox{NGC 4374}, these values are of the central dust disk and do not include the large-scale dust lane.\\
}
\end{deluxetable*}

Prior to early science observations with the Atacama Large Millimeter/submillimeter Array (ALMA), only a few ETGs were suitable candidates for BH mass measurements using other arrays. The previous generation of mm/sub-mm interferometers could resolve $r_{\rm g}$ for many nearby ETGs \citep{davis14}, but only a small percentage of these galaxies have detectable quantities of rapidly rotating cold gas in their nuclei. In the first successful case, \citet{dav13a} observed CO kinematics from within $r_{\rm g}$ and demonstrated the capacity of cold molecular gas to constrain BH masses. At its full capability, ALMA now delivers about an order of magnitude increase in sensitivity and angular resolution over previous arrays. Even in configurations with sub-km maximum baselines, this array is capable of routinely measuring gas kinematics within $\sim r_{\rm g}$ for many nearby ETGs.

In addition to resolving gas kinematics to pursue BH mass measurements, such ALMA observations are capable of probing molecular gas morphologies and chemistries, along with continuum properties, for a statistically significant number of early-type galaxy nuclei. Using, for example, resolved gas kinematics and derived radial mass profiles reveals whether the molecular gas disks are (at least formally) stable against gravitational fragmentation and the likelihood they host recent/ongoing star formation \citep{too64}. Observations of certain chemical tracers are expected to yield accurate column density measurements and estimates of gas-phase metallicities across the disks \citep{bay12}. Thermal continuum and CO isotopologue measurements will together place spatially resolved constraints on gas and dust temperatures, H$_2$ volume densities, gas-to-dust ratios, and the CO-to-H$_2$ conversion factor; the last of these is not well understood for ETGs or for the inner $\sim{\rm kpc}$ of galaxies in general \citep{san13}.

In ALMA \mbox{Cycle 2}, we began a program to determine sub-arcsecond scale ETG ${}^{12}$CO(2$-$1) morphologies and kinematics following a two-step process. First, we obtain sufficiently high resolution observations to detect high-velocity gas originating from within the BH sphere of influence. Then, when such gas is found, we conduct deeper, higher-resolution ALMA observations to map out the disk kinematics within $r_{\rm g}$ and accurately measure the BH masses. For our initial sample, we focused on ETGs with an (estimated) $r_{\rm g}\gtrsim 0\farcs3$. In this paper, we describe that sample of seven galaxies, characterize their CO(2$-$1) emission and kinematics, and present their resolved and unresolved continuum properties. Dynamical modeling for one galaxy, \mbox{NGC 1332}, has already been presented \citep{bar16b,bar16a}. In a future paper we will focus on modeling the CO(2$-$1) kinematics for the remainder of the sample.

\section{Sample Selection and Observations}
\label{obs}

\begin{figure*}[h!]
\begin{center}
\includegraphics[trim=0mm 0mm 0mm 0mm, clip, height=0.85\textheight]{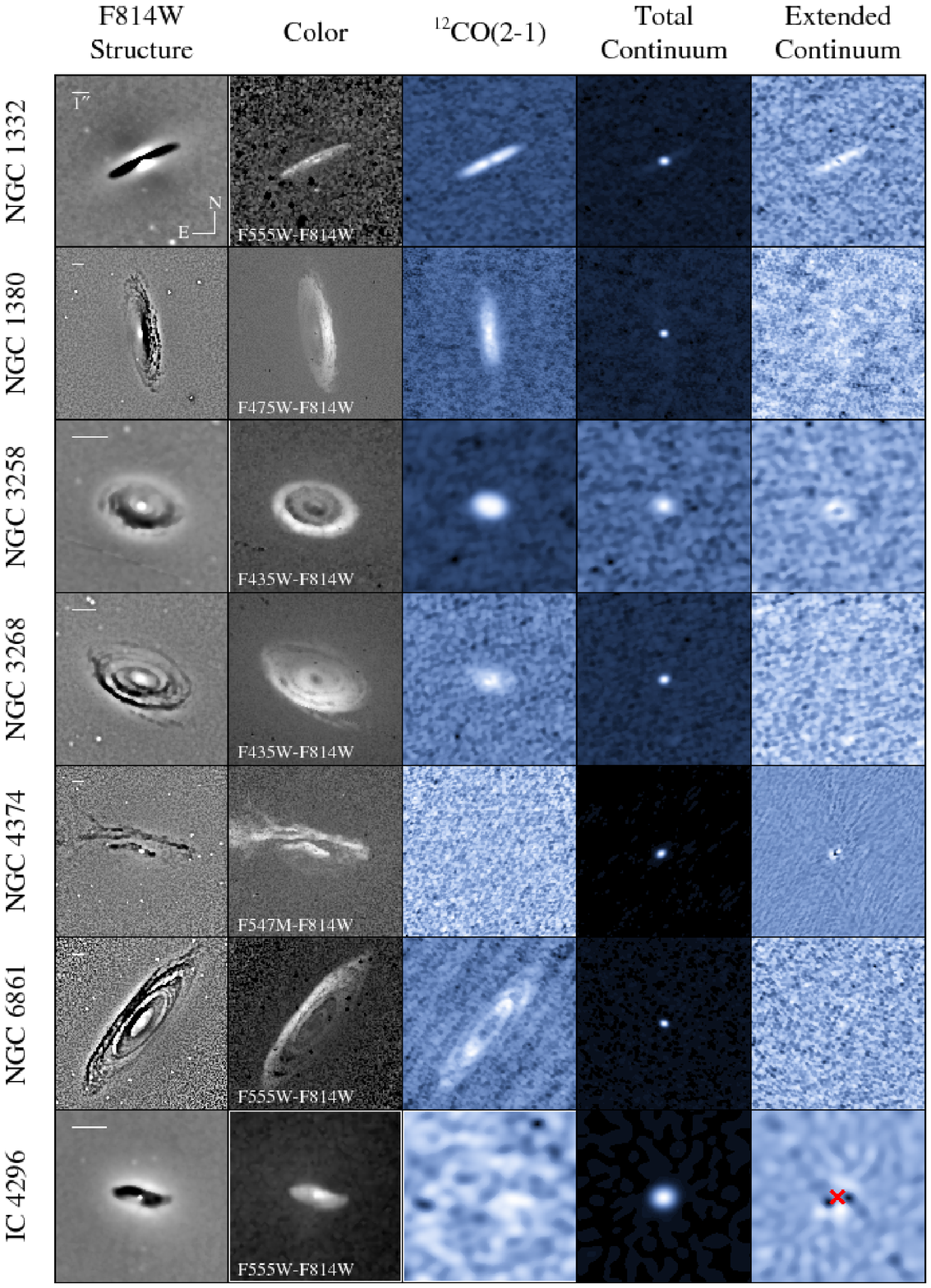}
\caption{Comparison of {\it HST} structure \citep{pog02} and two-band color maps with the CO(2-1) emission line and $\sim236$ GHz continuum images from ALMA Band 6 observations. The {\it HST} images are in linear gray scale while the ALMA images are displayed logarithmically. Bright regions in the structure maps indicate an excess of light, and darker regions in the color maps indicate bluer colors. The angular scale bar in each {\it HST} structure map represents $1\arcsec$ and applies to the subsequent ALMA images. These structure and color maps reveal regular dust disk morphology in all cases, and dust lanes for \mbox{NGC 4374} at radii past $r\gtrsim1\arcsec$. The CO(2$-$1) intensity maps are constructed by collapsing the ALMA data cubes over the velocity channels that show (or are expected to show) CO emission. Continuum maps are constructed by imaging the spectral windows over all channels that do not contain line emission. The extended continuum images are residual maps after modeling (and subtracting) the central peak continua in the {\it uv} plane. For \mbox{IC 4296}, point-source over-subtraction of the bright nucleus (at the location of the ``$\times$'' in the residual image) leads to the east-west imaging artifacts alongside its real extended continuum emission.}
\label{struct}
\end{center}
\end{figure*}

Targets for these \mbox{Cycle 2} observations were selected from ETG candidates with morphologically round dust disks as seen in {\it HST} images. \citet{ho02} demonstrated that the presence of circularly symmetric dust lanes in S0 and late-type galaxies tends to correlate with regular, symmetric ionized gas velocity fields, and we expect this trend to hold for their molecular gas kinematics. We imposed an additional selection criterion that the estimated angular $r_{\rm g}$ sizes should be large (here, $\gtrsim0\farcs3$) in projection along the major axis as $M_{\rm BH}$ measurements are easiest for galaxies having large projected $r_{\rm g}$. The candidates observed as part of Program 2013.1.00229.S were \mbox{NGC 1332}, \mbox{NGC 1380}, \mbox{NGC 3258}, \mbox{NGC 3268}, \mbox{NGC 6861} and \mbox{IC 4296}. In Program 2013.1.00828.S, ALMA observed \mbox{NGC 4374} (M84), while data for an additional object were not taken. These targets are luminous ($M_{K_s}<-24$ mag), nearby ($D\sim17$ to 47 Mpc; $z\sim0.0034$ to 0.012) galaxies whose closest large neighbors happen to lie $\gtrsim2\arcmin$ away in projection. Luminosity distances ($D_L$; see Table~\ref{tbl-obspars}) for these seven galaxies were derived using distance moduli measured by surface brightness fluctuations \citep{mei00,ton01}. The dust disks are very apparent in {\it HST} F814W structure maps (Figure~\ref{struct}), which highlight clumpy and filamentary dust \citep{pog02}. We also constructed two-band {\it HST} color maps (e.g., F555W$-$F814W; also Figure~\ref{struct}), which show more smoothly distributed dust. The \mbox{NGC 4374} color map reveals that, beyond $\sim1\arcsec$ from the nucleus, the dust is concentrated in an asymmetric dust lane. Despite its somewhat irregular dust morphology, we included \mbox{NGC 4374} in our sample as {\it HST} spectroscopy shows ionized gas in regular rotation within the central $0\farcs5$ (e.g., \citealp{bow97,wal10}), and ALMA offers the possibility of carrying out a direct comparison of circumnuclear CO and ionized gas kinematics.
 
Four of these ETGs have prior $M_{\rm BH}$ measurements. The \mbox{NGC 1332} BH mass was previously measured to be $(1.45\pm0.20)\times10^9$ $M_\sun$ using stellar dynamical modeling \citep{rus11}, while \citet{hum09} find a much lower mass of $(0.52^{+0.41}_{-0.28})\times 10^9$ $M_\sun$ using X-ray gas hydrostatic equilibrium constraints. The BH in \mbox{NGC 6861} was also measured with stellar dynamical modeling to have a mass of $(2.0\pm0.2)\times10^9$ $M_\sun$ \citep{rus13}. The BHs in \mbox{NGC 4374} and \mbox{IC 4296} were measured to have masses of $(9.25^{+0.98}_{-0.87})\times10^8$ $M_\sun$ \citep{wal10} and $(1.34^{+0.21}_{-0.19})\times10^9$ $M_\sun$ \citep{dal09}, respectively, using ionized gas dynamical modeling. In the first BH mass measurement to resolve $r_{\rm g}$ with ALMA, \citet{bar16a} model high-resolution ($0\farcs044$) \mbox{Cycle 3} \mbox{NGC 1332} CO(2$-$1) observations and determine its BH mass to be $(6.64^{+0.65}_{-0.63})\times10^8$ $M_\sun$. Based solely on averages over several velocity dispersion measurements in the HyperLeda catalog (\citealp{mak14}; see Table\ref{tbl-obspars}) and the $M_{\rm BH}-\sigma_\star$ relationship \citep{kor13}, the remaining three ETGs -- \mbox{NGC 1380}, \mbox{NGC 3258}, and \mbox{NGC 3268} -- have expected BH masses of $(3.9,9.3,6.3)\times10^8$ $M_\sun$, respectively. The projected $r_{\rm g}$ for our sample range between $\sim0\farcs25$ and $0\farcs61$ and should be at least nearly resolved with the ALMA observations presented here.

Our ALMA Band 6 observations consisted of single pointings with three $\sim2$ GHz bandwidth spectral windows, the first centered on the redshifted ${}^{12}{\rm CO}(2-1)$ 230.538 GHz line and the remaining two measuring the continuum at observed frequencies of $\sim226$ and $\sim245$ GHz. Each target was observed with channel widths of 488 kHz and 15.6 MHz (corresponding to velocity channel widths of 0.62 and 19.8 km s$^{-1}$) for the line and continuum spectral windows, respectively. The observations were flux calibrated using Titan, Ceres, and Ganymede when available, along with J0334$-$4008 and J2056$-$4719 from the ALMA quasar catalog when no solar system standards were nearby. These flux standards combine to give $\lesssim10\%$ flux scale uncertainty at any time \citep{fom14}. We therefore propagate a 10\% uncertainty into all flux and flux density measurements. Data were processed using versions 4.2.2 and 4.5 of the Common Astronomy Software Application (CASA; \citealp{mcm07}) and version 4.2.2 of the standard ALMA pipeline. Image deconvolution was performed using the CASA CLEAN task. As each target possesses a nuclear continuum source, we applied continuum phase self-calibration to each data set with the exception of \mbox{NGC 3258}, whose nuclear continuum dynamical range did not improve with self-calibration. We also applied amplitude self-calibration to the data sets of the two brightest continuum sources, \mbox{NGC 4374} and \mbox{IC 4296}. On-source integration times ranged from 20 to 50 minutes, with typical rms noise levels of $\sim400$ $\mu$Jy beam$^{-1}$ per 20 km s$^{-1}$ channel, and $\sim40$ $\mu$Jy beam$^{-1}$ after imaging of the continuum spectral window data into a single image. The angular resolution of these observations range between $\sim0\farcs2$ and $0\farcs5$ (see Table~\ref{tbl-obsalma}), corresponding to an average projected spatial resolution of $\sim50$ pc.

\begin{deluxetable*}{lcccccc}[h!]
\tabletypesize{\scriptsize}
\tablecaption{ALMA Observational Parameters\label{tbl-obsalma}}
\tablewidth{0pt}
\tablehead{
\multirow{2}{*}{\phantom{m}Galaxy} & \multicolumn{1}{c}{Obs.} & \multicolumn{1}{c}{$t_{\rm obs}$} & \multicolumn{1}{c}{Baseline (m)} & \colhead{$b_{\rm maj}$ / $b_{\rm min}$} & \colhead{Freq. Range / Bandwidth} & \colhead{Self-} \\
 & \multicolumn{1}{c}{Date} & \multicolumn{1}{c}{(min)} & \multicolumn{1}{c}{min / max} & $('')$ & \colhead{(GHz)} & \colhead{Calibration} \\
\multicolumn{1}{c}{(1)} & (2) & \multicolumn{1}{c}{(3)} & \multicolumn{1}{c}{(4)} & (5) & \multicolumn{1}{c}{(6)} & (7)
 }
\startdata
NGC 1332 & 2014-09-01 & 22.3 & 33 / 1091 & 0.32 / 0.23 & $226.4-246.1$ / 5.9 & p \\
\rule{0pt}{4ex}\multirow{2}{*}{NGC 1380} & 2015-06-11 & \multirow{2}{*}{22.8} & 21 / 783\phantom{ }\phantom{ } & \multirow{2}{*}{0.24 / 0.18} & \multirow{2}{*}{$226.2-245.9$ / 5.9} & \multirow{2}{*}{p} \\
 & 2015-09-18 &  & 41 / 2125 &  &  & \\
\rule{0pt}{4ex}\multirow{2}{*}{NGC 3258} & 2014-07-21 & \multirow{2}{*}{22.4} & 18 / 784\phantom{ }\phantom{ } &  \multirow{2}{*}{0.48 / 0.40} & \multirow{2}{*}{$225.5-245.1$ / 5.9} & \multirow{2}{*}{$-$} \\
 & 2015-06-12 &  & 21 / 784\phantom{ }\phantom{ } &  &  & \\
\rule{0pt}{4ex}\multirow{2}{*}{NGC 3268} & 2014-07-21 & \multirow{2}{*}{22.3} & 18 / 784\phantom{ }\phantom{ } & \multirow{2}{*}{0.45 / 0.40} & \multirow{2}{*}{$225.5-245.1$ / 5.9} & \multirow{2}{*}{p} \\
 & 2015-06-12 &  & 21 / 784\phantom{ }\phantom{ } &  &  & \\
\rule{0pt}{4ex}NGC 4374 & 2015-08-16 & 50.9 & 43 / 1574 & 0.35 / 0.26 & $227.2-246.6$ / 7.9 & ap \\
\rule{0pt}{4ex}NGC 6861 & 2014-09-01 & 23.9 & 34 / 1091 &  0.32 / 0.23 & $225.5-245.1$ / 5.9 & p \\
\rule{0pt}{4ex}IC 4296 & 2014-07-22 & 29.6 & 18 / 780\phantom{ }\phantom{ } &  0.52 / 0.43 & $224.8-244.4$ / 5.9 & ap \\
\enddata
\tablecomments{Properties of the ALMA \mbox{Cycle 2} observations. Col. (3): On-source integration time. Col. (4): Minimum and maximum baselines of the specific array configurations. Col. (5): FWHMs of the synthesized beam major and minor axes when using natural weighting for CO-faint galaxies \mbox{NGC 4374} and \mbox{IC 4296}, and Briggs weighting ($r=0.5$) for the remainder. Col. (6): Frequency range including the continuum basebands, followed by the combined bandwidth of all continuum spectral windows. Col. (7): Indicates phase-only (p) or amplitude and phase (ap) continuum self-calibration. \\
}
\end{deluxetable*}

The continuum of each measurement set was imaged using both natural weighting for the greatest sensitivity to faint, extended emission, and Briggs weighting \citep{bri95} with robustness parameter $r=0.5$ for a good trade-off between resolution and sensitivity. We further explored the frequency dependence of the continuum by imaging the visibilities into multiple, separate continuum-only spectral window images. To isolate the molecular emission, the continuum contributions were first removed in the {\it uv} plane from the spectral window containing the CO(2$-$1) line. The resultant continuum-subtracted visibilities were then imaged into data cubes using Briggs weighting ($r=0.5$) with 20 km s$^{-1}$ channels if we observe high S/N CO emission, and 40 km s$^{-1}$ channels using natural weighting otherwise. Velocities were calculated in the barycentric frame using the optical definition of radial velocity. Spatial pixel sizes varied for each target and were chosen to adequately sample the beam minor axis.

\section{CO(2$-$1) Line}
\label{coline}

Inspection of the continuum-subtracted image cubes reveals significant CO(2$-$1) line emission across many channels for \mbox{NGC 1332}, \mbox{NGC 1380}, \mbox{NGC 3258}, \mbox{NGC 3268}, and \mbox{NGC 6861}. We also detect both CO(2$-$1) emission (at low S/N) and absorption in \mbox{NGC 4374} and \mbox{IC 4296}. Figure~\ref{struct} shows integrated intensity maps, made by summing the image cubes over the channel ranges that show line emission. For the two galaxies that have faint CO emission, the intensity maps show the cubes collapsed over $\pm400$ km s$^{-1}$ from the systemic velocity ($v_{\rm sys}$). Within the primary beam half-power width ($\sim25\arcsec$), no CO line emission is detected outside these dust disk regions. In the next section, we discuss the emission properties of the CO-bright subsample, and in $\S$\ref{co-faint} we investigate the emission and absorption characteristics of the two CO-faint targets.

\subsection{CO-Bright Galaxies}
\label{co-bright}

\begin{figure*}
\begin{center}
\includegraphics[trim=0mm 0mm 0mm 0mm, clip, width=0.95\textwidth]{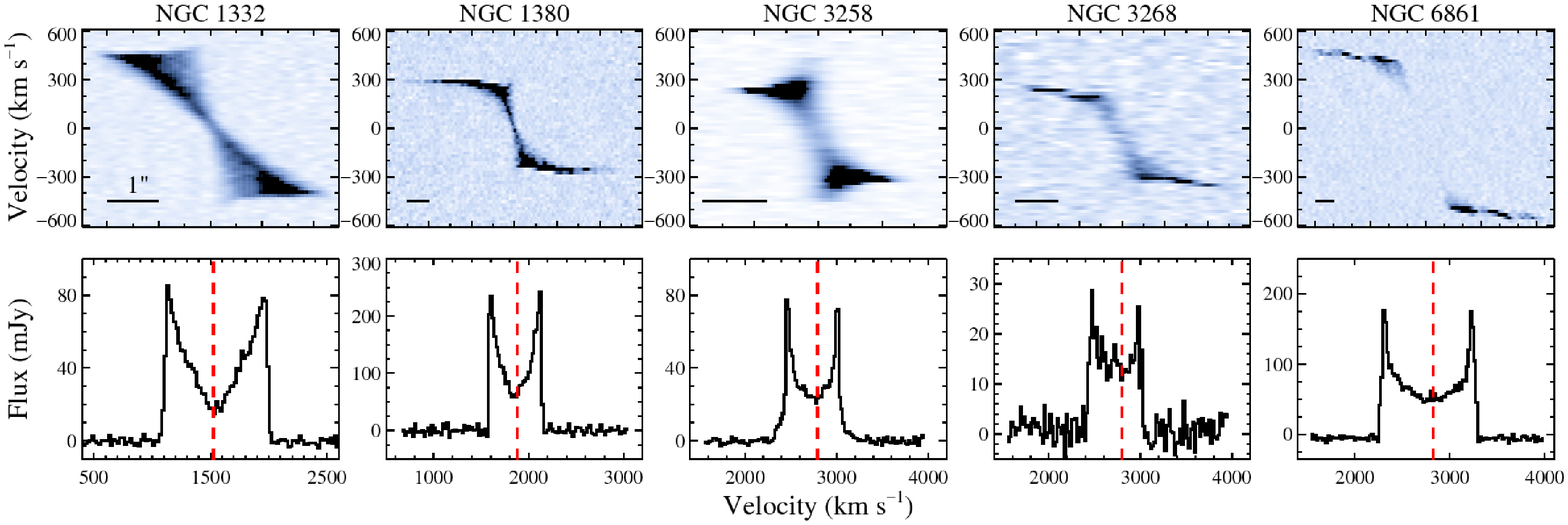}
\caption{{\it Top}: CO(2$-$1) position-velocity diagrams (PVDs) for the CO-bright galaxies, extracted with a width equal to the synthesized beam FWHM projected onto the disk major axis. To bring out faint emission, we display these PVDs using a negative, arcsinh color scale. Central velocity upturns are observed in \mbox{NGC 1332} and \mbox{NGC 3258}, with the possible addition of \mbox{NGC 1380}. {\it Bottom}: Velocity profiles of these targets, formed by spatially integrating over the optical dust disk regions for each 20 km s$^{-1}$ wide channel. The dashed lines indicate the host galaxy $v_{\rm sys}$, as reported in the NASA/IPAC Extragalactic Database (NED).}
\label{pvd_velchan}
\end{center}
\end{figure*}

The bright CO(2$-$1) emission seen in \mbox{NGC 1332}, \mbox{NGC 1380}, \mbox{NGC 3258}, \mbox{NGC 3268}, and \mbox{NGC 6861} directly coincides with optically thick dust disks as seen in the structure maps of the corresponding optical {\it HST} images. Inspecting their image cubes reveals clear rotation about the nuclear continuum sources. To visualize their rotational properties, we integrated the flux densities in each channel over the elliptical regions coinciding with the optical dust disks. These CO velocity profiles (Figure~\ref{pvd_velchan}) exhibit nearly symmetric double-horned features characteristic of rotating disks with emission ranging between $\pm300$ and $\pm500$ km s$^{-1}$ from $v_{\rm sys}$. We also constructed a position-velocity diagram (PVD; also shown in Figure~\ref{pvd_velchan}) for each galaxy by extracting a slice along the major axis as determined from the average disk position angle (measured in $\S$\ref{kin}). The extraction widths were set to the projected CLEAN beam FWHM along the disk major axis. A central rise in the PVD envelope to high velocity at small radius, which indicates a massive, compact object at the disk center, is unambiguously observed only in \mbox{NGC 1332} and \mbox{NGC 3258}, with a marginal detection in \mbox{NGC 1380} within the central $\sim0\farcs1$. The PVD for \mbox{NGC 6861} shows a $\sim2\arcsec-$wide hole in CO emission.

For a disk whose line emission primarily originates within the sphere of influence of a massive object, the double-horned peaks in its velocity profile suggest near-perfect Keplerian rotation. This is not the case for the roughly kpc-scale disks in our CO-bright sample. Comparing the velocity profiles to the PVDs, we find that most of the contributions to the peaks of the double-horned profiles originate far ($\gtrsim100$ pc) from the disk centers, at which radii the extended stellar mass profiles dominate over the BH mass. At velocities beyond these peaks (especially for \mbox{NGC 1380}, \mbox{NGC 3268}, and \mbox{NGC 6861}), the sharp decline in integrated flux density is the result of flat (or nearly flat) rotation curves in the outer disk regions. The Keplerian velocity signatures detected in the PVDs of \mbox{NGC 1332} and \mbox{NGC 3258} appear only as faint, high-velocity wings in their respective velocity profiles.

CO surveys conducted with single-dish telescopes generally cannot resolve circumnuclear gas disk rotation within nearby galaxies, although velocity profiles from these observations can determine line luminosities (or upper limits) and velocity ranges for a substantial number of ETGs. Such surveys have shown, for instance, that the CO detection rate may be slightly environment-dependent (i.e., field vs. cluster; \citealp{you11}) and that the CO emission velocity width strongly correlates with the host galaxy mass (the CO Tully-Fisher relationship; \citealp{ho07,dav11a}). However, the detection of a double-horned profile in low angular resolution observations does not necessarily indicate that the galaxy will be a good candidate for $M_{\rm BH}$ measurement. These CO-bright PVDs demonstrate that the line emission originating from within $r_{\rm g}$ is generally faint, and is not detected beyond a deprojected velocity of $\sim600$ km s$^{-1}$ for these five galaxies. In the absence of strong, broad ($\sim1000$ km s$^{-1}$ from $v_{\rm sys}$) velocity wings, spatially resolved data are necessary to confidently detect the BH kinematic signature. Based on our \mbox{Cycle 2} PVDs, interferometric observations with a spatial resolution that corresponds to $\sim r_{\rm g}$ appear to be sufficient to identify high-velocity, central gas emission.

\begin{figure*}[h!]
\begin{center}
\includegraphics[trim=0mm 0mm 0mm 0mm, clip, width=0.975\textwidth]{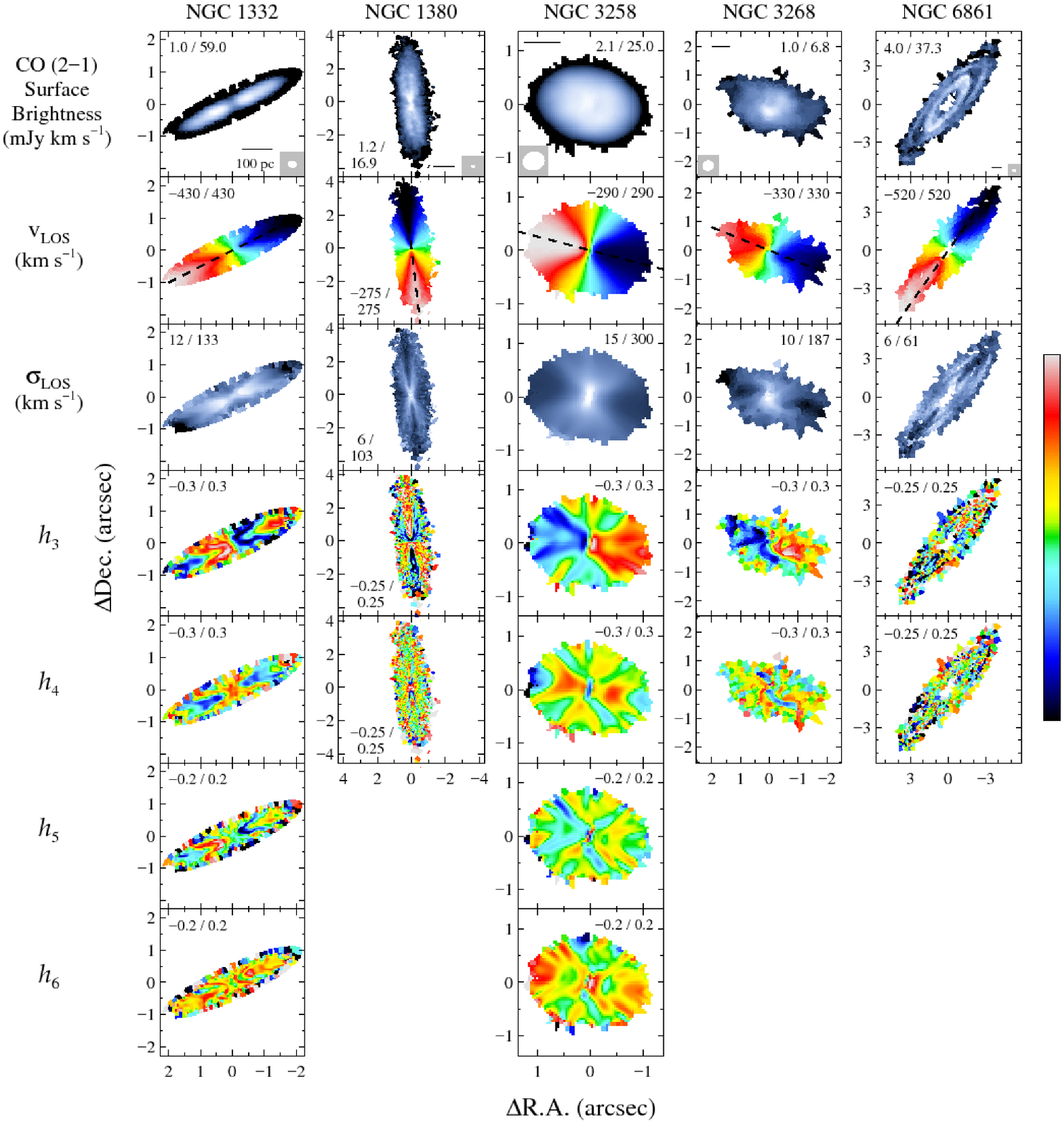}
\caption{CO(2$-$1) line parameters for the CO-bright galaxies. Due to line complexity, we parametrized the line profiles using Gauss-Hermite (GH) polynomials. Shown are the CO(2$-$1) surface brightness maps, along with maps for the line-of-sight velocity $v_{\rm LOS}$ and dispersion $\sigma_{\rm LOS}$, and GH moments from $h_3$ up to $h_6$. The galaxy systemic velocities have been removed from $v_{\rm LOS}$. The scale bar in each CO flux map denotes a physical 100 pc, and alongside is shown the beam size. At the disk edges (largely undisplayed) and near the center of \mbox{NGC 6861}, the individual spectra were Voronoi binned spatially to achieve sufficient S/N to fit the line profiles. For each target, its kinematic maps are displayed down to the flux limit where its $v_{\rm LOS}$ field remains well behaved. The values provided alongside the parameter maps are the minimum and maximum data values that correspond to the ranges in intensity or color. The dashed lines superimposed on each $v_{\rm LOS}$ profile are the stellar photometric major axes position angles measured from {\it HST} images.}
\label{ghmaps}
\end{center}
\end{figure*}

\subsubsection{Line Profile Fitting}
\label{line_profile}

To characterize the velocity fields of the five CO-bright disks, we parametrize the line profiles at each spatial pixel with Gauss-Hermite (GH) profiles \citep{mar93}, including the flux, Gaussian line-of-sight (LOS) velocity $v_{\rm LOS}$ and dispersion $\sigma_{\rm LOS}$, and higher-order terms ($h_3$, $h_4$, ...) that account for symmetric (even terms) and anti-symmetric (odd) deviations from a purely Gaussian profile. The $\sim0\farcs3-$resolution beam size entangles CO emission across large ($\sim50$ pc) physical regions and, especially near the nucleus, emission from a wide range of intrinsic velocities becomes spectrally and spatially blended. Within their central beam areas, the PVDs show broad velocity envelopes with complex and asymmetric line profiles (see \citealp{bar16b}). We use $h_3$ and $h_4$ components in all spectral fits, and include $h_5$ and $h_6$ when modeling the more complicated and higher S/N line profiles in \mbox{NGC 1332} and \mbox{NGC 3258} for which $h_5$ and $h_6$ remain large beyond the central beam area. To measure line parameters with higher precision, adjacent spectra are combined together based on their line S/N using the Voronoi binning method \citep{cap03}. This spatial binning primarily affects regions near the edge of the CO-emitting disks, although it also proves useful for the central $2\arcsec$ of \mbox{NGC 6861}. There are no {\it a priori} limits on these GH coefficients, especially given the level of beam smearing. We fix limits of $|h_3,h_4|<0.35$ (and $|h_5,h_6|<0.25$ when fitting the \mbox{NGC 1332} and \mbox{NGC 3258} emission) that allows for reasonable line profile fits while avoiding model line profiles with strong double-peaked features. We determine the formal uncertainty of each GH term by adding random Gaussian noise to each Voronoi-binned spectrum with dispersion equal to the line residual standard deviation of its GH profile fit. The resampled spectrum is again fit with a GH profile to measure new line parameters; this process is repeated 500 times for each binned spectrum, and we set the formal uncertainties of each coefficient to the standard deviation of all its resampled fit values.

Maps of the CO(2$-$1) line flux, $v_{\rm LOS}$ and $\sigma_{\rm LOS}$, and GH coefficients are shown for each galaxy in Figure~\ref{ghmaps}. Line flux in the observed frame (in Jy km s$^{-1}$) for each binned spectrum is determined by integrating the profile between $v_{\rm LOS}\pm3\sigma_{\rm LOS}$. The CO(2$-$1) flux maps are spatially coincident with, and similar in shape to, the optical dust disks. Flux maps indicate significant deviations from smooth emission only for two of the most highly resolved disks -- \mbox{NGC 1380} and \mbox{NGC 6861} -- where the synthesized beams are far smaller than the disk sizes. \mbox{Cycle 3} observations of the \mbox{NGC 1332} molecular disk at $\sim0\farcs044$ resolution \citep{bar16a} do show significantly more CO(2$-$1) substructure than the \mbox{Cycle 2} flux map, suggesting that clumpy emission is standard for these circumnuclear disks.

\begin{figure*}[ht!]
\begin{center}
\includegraphics[trim=0mm 0mm 0mm 0mm, clip, width=0.975\textwidth]{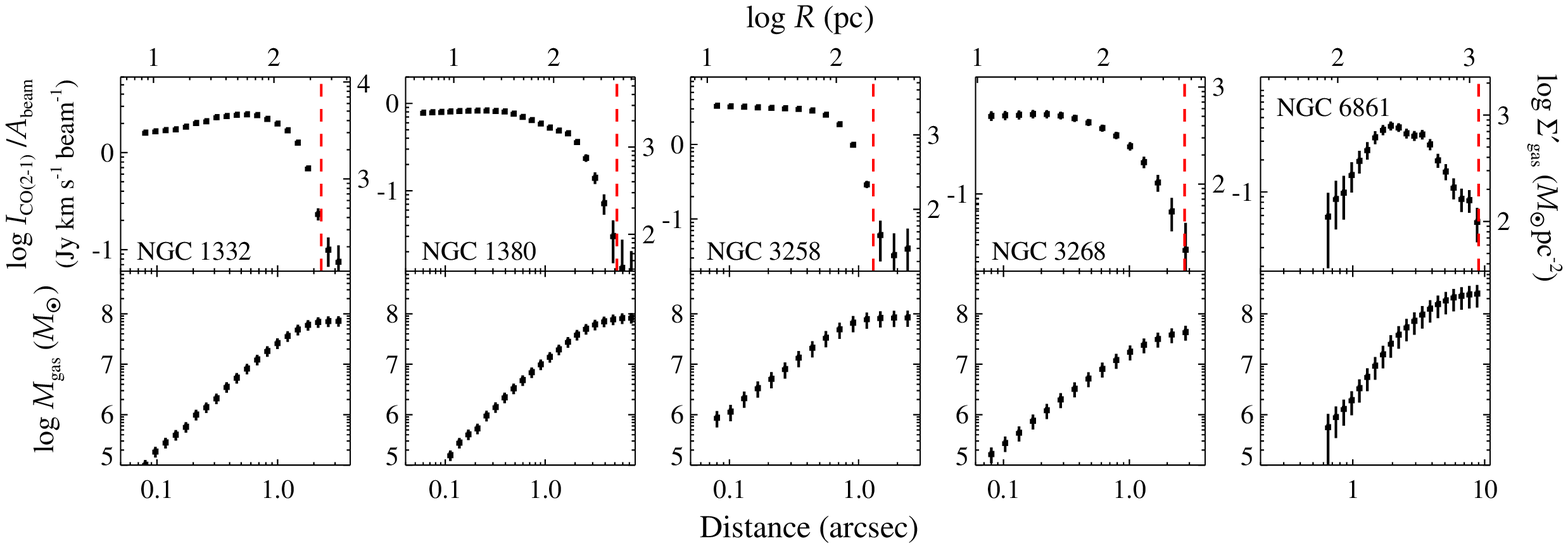}
\caption{{\it Top}: Radial CO(2$-$1) surface brightness profiles, averaged on elliptical annuli using average position angles and axis ratios from Table~\ref{tbl-kin}. The right vertical axes indicate the corresponding projected surface mass densities in units of $M_\sun$ pc$^{-2}$. The dashed lines show the radial extent of the optically opaque dust features in the {\it HST} images. {\it Bottom}: Cumulative gas mass profiles assuming $\alpha_{\rm CO}=3.1$ $M_\sun$ pc$^{-2}$ (K km s$^{-1}$)$^{-1}$ and incorporating the expected helium contributions.}
\label{massprof}
\end{center}
\end{figure*}

The molecular gas kinematics of the five CO-bright galaxies are dominated by regular, disk-like rotation. We characterize the deviations from coplanar, circular rotation in $\S$\ref{kin}. Visual inspection shows moderate kinematic warping in \mbox{NGC 3268}; for the remainder of this subsample the disk kinematic axes generally conform to the larger-scale stellar photometric axes. Stellar photometric major axis position angles (PAs), measured from the inner several kpc of {\it HST} images, agree to within $\lesssim5\degr$ with the average CO kinematic axes. This does not necessarily mean that the molecular gas is aligned with the stellar kinematics, since mismatches between {\it stellar} photometric and kinematic axes can exist in ETGs. However, \citet{kra11} find that the stellar photometric and kinematic axes nearly always agree to within $\sim 10\degr$ for fast-rotating ETGs. In addition, luminous ($M_{K_s}<-24$ mag) ETGs with molecular (and/or \ion{H}{1}, \ion{H}{2}) gas nearly always show kinematic alignment between stellar and gaseous rotation \citep{dav11b}. Previously published integral field unit (IFU) spectroscopy of  \mbox{NGC 1332} \citep{rus11}, \mbox{NGC 1380} \citep{ric14}, and \mbox{NGC 6861} \citep{rus13} shows that their molecular gas disks co-rotate (to an accuracy of $\lesssim5\degr$) with respect to their stars. Spatially resolved stellar kinematics are not available for \mbox{NGC 3258} and \mbox{NGC 3268}, although based on the results for the three other CO-bright galaxies we expect that their molecular disks also co-rotate with the stars.

The line dispersions generally increase towards the disk kinematic centers due to beam smearing. The exceptions are \mbox{NGC 1332}, whose $\sigma_{\rm LOS}$ diminishes from $\sim140$ to $\sim100$ km s$^{-1}$ in the inner $0\farcs25$, and \mbox{NGC 6861}, which possesses a wide hole in the CO emission. X-shaped features appear in all five $\sigma_{\rm LOS}$ maps, and the high CO line widths forming these features are due to beam smearing of large velocity gradients at these locations. These gradients are lowest along the major axes near the disk edges, where the observed line dispersions fall to between 6 and 15 km s$^{-1}$ with typical uncertainties of $\sim1$ km s$^{-1}$. For comparison, the gas sound speed $c_s$ is expected to be $\sqrt{\gamma_d k_B T_d/(\mu m_p)}\sim0.3$ km s$^{-1}$ for an assumed $T_d\sim30$ K (typical for molecular gas in other ETGs; \citealp{bay13}), specific heat $\gamma_d\sim 1$, and mean molecular mass $\mu\sim2.5$ for the predominantly H$_2+$He disks. For \mbox{NGC 1380} and \mbox{NGC 6861} in particular, their minimum line widths ($\sigma_{\rm LOS}\sim6$ km s$^{-1}$) are much smaller than the 20 km s$^{-1}$ channel widths. We tested the accuracy of our line width fits by re-imaging the \mbox{NGC 1380} and \mbox{NGC 6861} visibilities into 4 km s$^{-1}$ channels; line profile fits to these more finely sampled data cubes show line widths that agree to within $\sim1$ km s$^{-1}$ with those measured in cubes with coarser (20 km s$^{-1}$) velocity binning. Even though the effect is lowest near the disk edges and along the major axes, portions of the lowest observed $\sigma_{\rm LOS}$ values are still the result of beam smearing. In a future paper, we will describe gas dynamical modeling to constrain the intrinsic gas line widths as a function of radius.

\subsubsection{Surface Brightness and Mass Profiles}
\label{co_sb}

For the CO-bright galaxies, line intensities $I_{\rm CO(2-1)}$ integrated over the emitting areas range between $\sim$10 and 100 Jy km s$^{-1}$ (Table~\ref{tbl-linemeas}). The H$_2$ mass is determined by computing $M_{\rm H_2}=\alpha_{\rm CO}L^\prime_{\rm CO}$ and the CO luminosity $L^\prime_{\rm CO}$ is related to observables \citep{car13} by:
\begin{equation}
\label{eq-lp}
L^\prime_{\rm CO}=3.25\times10^7 S_{\rm CO}\Delta v \frac{D_L^2}{(1+z)^3 \nu_{\rm obs}^2}\;{\rm K\; km\; s^{-1}\; pc^2}\, .
\end{equation}
The line intensity is $S_{\rm CO}\Delta v$ (typically of the $J=1-0$ transition) and $D_L$ is the galaxy luminosity distance. We convert our integrated CO(2$-$1) flux measurements into H$_2$ masses using average $R_{21}$ and $\alpha_{\rm CO}$ values from a sample of nearby, late-type galaxies \citep{san13}, assuming $R_{21}=I_{\rm CO(2-1)}/I_{\rm CO(1-0)}\approx0.7$ (corresponding to an excitation temperature $T_{\rm ex}\sim5-10$ K; e.g., \citealp{lav99}) to transform the CO(2$-$1) line intensities into estimated CO($1-0$) values, and using $\alpha_{\rm CO}=3.1$ $M_\sun$ pc$^{-2}$ (K km s$^{-1}$)$^{-1}$ as the extragalactic mass-to-luminosity ratio. The average molecular hydrogen mass for the CO-bright targets is then $\sim10^8$ $M_\sun$, although the most appropriate CO-to-H$_2$ conversion factor for our sample (or ETGs in general) is not currently known. For comparison, the stellar mass within the central 100 pc measured from the {\it HST} F814W images (assuming an $I-$band mass-to-light ratio $\Upsilon_{I}\approx5$ $M_\sun/L_\sun$; e.g., \citealp{wal12}) is on the order of $10^9$ $M_\sun$. In addition to substantial scatter (0.3 dex) about the average $\alpha_{\rm CO}$ for extragalactic sources, \citet{san13} find that this mass-to-light ratio typically decreases by a factor of $\sim2$ (and in some cases by nearly an order of magnitude) in the central kpc of late-type galaxies. Given the large scatter in $\alpha_{\rm CO}$, the derived gas masses are subject to potentially large systematic uncertainties. Measuring and modeling CO spectral line energy distributions (SLEDs) for these galaxies should allow a more precise $\alpha_{\rm CO}$ determination (even using only spatially resolved, low$-J$ lines observable with ALMA; e.g., \citealp{sai17}).

\begin{deluxetable*}{lcccccccc}[ht!]
\tabletypesize{\scriptsize}
\tablecaption{{}$^{12}$CO(2$-$1) Disk Parameters\label{tbl-linemeas}}
\tablewidth{0pt}
\tablehead{
\multirow{2}{*}{\phantom{m}Galaxy} & \colhead{RMS Noise} & \colhead{$R_{\rm d}$} & \colhead{$I_{\rm CO(2-1)}$} & \colhead{$\Delta v_{\rm CO(2-1)}$} & \colhead{$W_{50}$} & \colhead{$L^\prime_{\rm CO(2-1)}$} & \colhead{$M_{\rm gas}$} & \colhead{$N_{\rm H_2}$} \\
 & \colhead{(mJy beam$^{-1}$)} & \colhead{(pc)} & \colhead{(Jy km s$^{-1}$)} & \colhead{(km s$^{-1}$)} & \colhead{(km s$^{-1}$)} & \colhead{($10^6$ K km s$^{-1}$ pc$^{-2}$)} & \colhead{($10^7$ $M_\sun$)} & \colhead{($10^{21}$ cm$^{-2}$)} \\
\multicolumn{1}{c}{(1)} & (2) & (3) & (4) & (5) & (6) & (7) & (8) & (9)
 }
\startdata
NGC 1332 & 0.40 & 256 & 39.94 (3.99) & 1000 & 880 & \phantom{0}12.1 (2.3) & \phantom{00}7.3 (1.4) & 71 \\
\rule{0pt}{4ex}NGC 1380 & 0.39 & 426 & 78.35 (7.84) & 580 & 560 & \phantom{0}14.0 (2.7) & \phantom{00}8.4 (1.6) & 37 \\
\rule{0pt}{4ex}NGC 3258 & 0.35 & 164 & 23.89 (2.39) & 880 & 580 & \phantom{0}14.1 (3.8) & \phantom{00}8.5 (2.3) & 29 \\
\rule{0pt}{4ex}NGC 3268 & 0.45 & 373 & 10.73 (1.07) & 620 & 580 & \phantom{00}7.6 (1.9) & \phantom{00}4.6 (1.2) & 11 \\
\rule{0pt}{4ex}NGC 4374 & 0.47 & -- & \phantom{0}4.81 (0.70) & 750 & -- & 0.94 (0.17) & 0.57 (0.10) & 4.0 \\
\rule{0pt}{4ex}NGC 6861 & 0.62 & 784 & 93.93 (9.39) & 1040 & 980 & 42.6 (14.8) & \phantom{0}25.6 (8.9) & 19 \\
\rule{0pt}{4ex}IC 4296 & 0.41 & -- &  \phantom{0}0.76 (0.18) & 600 & -- & 1.03 (0.30) & 0.62 (0.18) & 9.8 \\
\enddata
\tablecomments{Molecular gas properties from CO(2$-$1) flux measurements. Col. (2): The rms background noise per 20 km s$^{-1}$ channel of the Briggs weighted ($r=0.5$) image cube. For \mbox{NGC 4374} and \mbox{IC 4296}, these noise levels are per 40 km s$^{-1}$ channel in naturally-weighted image cubes. Col. (3): The intrinsic radial extent of the CO emission from surface brightness measurements. Col. (4): Integrated CO(2$-$1) flux measurements over the entire dust disks. For \mbox{NGC 4374}, this $I_{\rm CO(2-1)}$ measurement only integrates over the inner ($\sim1\arcsec$ radius) dust disk. Col. (5): Velocity range over which CO(2$-$1) emission is detected. Col. (6): Width of the CO velocity profile at 50\% of the peak intensity. Col. (7): Integrated line surface brightness temperature. Col. (8): Integrated gas mass of the molecular disk, assuming $\alpha_{\rm CO}=3.1$ $M_\sun$ pc$^{-2}$ (K km s$^{-1}$)$^{-1}$, a CO(2$-$1)/CO(1$-$0)$\approx0.7$ line ratio, and including contributions from helium. We give ($1\sigma$) uncertainties on $I_{\rm CO(2-1)}$, $L^\prime{}_{\rm CO(2-1)}$, and $M_{\rm gas}$ in parentheses; these include a 10\% uncertainty in the absolute flux scaling, and we also propagate uncertainties in $D_L$ into the $L^\prime{}_{\rm CO(2-1)}$ and $M_{\rm gas}$ confidence ranges. Col. (9): Estimated H$_2$ column densities made by averaging these $M_{\rm gas}$ masses over the CO disk radius $R_{\rm d}$ or (for the CO-faint targets) their central dust disk areas. \\
}
\end{deluxetable*}

The total atomic hydrogen mass of late-type galaxies often dominates the total molecular gas mass (e.g., \citealp{big08}), while in ETGs the total $M_{\rm H_2}\gtrsim M_{\rm H\,I}$ (see \citealp{ser12,ala13}). Only three galaxies in our sample have \ion{H}{1} detections or upper limits based on 21-cm observations -- \mbox{NGC 1332} with $\log(\,M_{\rm H\,I}/M_\sun)\sim8.8$ \citep{bal79}, \mbox{NGC 1380} with $\log(\,M_{\rm H\,I}/M_\sun)<8.6$ \citep{huc89}, and \mbox{NGC 4374} with $\log(\,M_{\rm H\,I}/M_\sun)\sim9.3$ \citep{dav73}. Beam sizes for these \ion{H}{1} observations are several arcminutes (corresponding to several tens of kpc), and neutral hydrogen envelopes tend to be much more extended than the molecular gas profiles. For late-type galaxies, \citet{ler08} and \citet{big08} find that the \ion{H}{1} surface mass density $\Sigma_{\rm H\,I}$ saturates at $\sim10$ $M_\sun$ pc$^{-2}$ and that molecular contributions dominate when $\Sigma_{\rm H_2+H\,I}\gtrsim14$ $M_\sun$ pc$^{-2}$. Assuming that any neutral hydrogen is also in a disk-like configuration and this \ion{H}{1} saturation limit applies to the $\sim 500$ pc radii circumnuclear gas disks in our ETG sample, neutral hydrogen should contribute at most $\sim10^7$ $M_\sun$ to the gas mass $M_{\rm gas}$ over the CO-emitting regions. Helium is included in the total gas mass as $M_{\rm gas}=M_{\rm H_2}(1+f_{\rm He})$ where $f_{\rm He}=0.36$ is the estimated mass-weighted helium fraction.

Our five CO-bright galaxies have $M_{\rm gas}$ values that range between $(0.5-2.6)\times10^8$ $M_\sun$, with a median of a little under $10^8$ $M_\sun$ (see Table~\ref{tbl-linemeas}). These results are similar to the interferometric ($\sim4\arcsec$ beam) gas measurements from a subsample of the ATLAS$^{\rm 3D}$ ETG survey, which finds $M_{\rm H_2}$ in the range of $(0.6-21.4)\times10^8$ $M_\sun$ for their CO-detected galaxies (using an $\alpha_{\rm CO}\approx6.5$; \citealp{ala13}), and those with disk-like gas morphology and rotation have a median $M_{\rm H_2}\approx 4.5\times10^8$ $M_\sun$. After adjusting for different $\alpha_{\rm CO}$ values, the ATLAS$^{\rm 3D}$ galaxies with interferometric data have a median H$_2$ mass that is four times larger than the median of our $M_{\rm H_2}$ values. This discrepancy is likely the result of ATLAS$^{\rm 3D}$ interferometric targets being selected from the brightest and largest CO sources in a full single-dish survey \citep{you11}.

Projected surface brightness and mass profiles (Figure~\ref{massprof}) have been created by averaging the CO flux in elliptical annuli centered on the continuum peaks; we fix the ellipse major axis PAs and axis ratios to the average kinemetry fit values as described in $\S$\ref{kin}. While most of the CO profiles peak at the disk centers, \mbox{NGC 1332} and \mbox{NGC 6861} reach their maximal surface brightnesses at $0\farcs5$ and $1\farcs9$ from rotation centers, respectively. The intrinsic radial size $R_{\rm d}$ of these disks is estimated by $R_{\rm d}^2=R^\prime{}_{\rm d}^2-\sigma_{\rm beam,CO}^2$ where $R^\prime{}_{\rm d}$ is the observed radial CO(2$-$1) extent and $\sigma_{\rm beam,CO}$ is the Gaussian dispersion size of the synthesized beam projected along the disk major axis. The intrinsic disk sizes measured using CO emission are equivalent (within $\sim20\%$) to those inferred from the largest angular extent of the optical dust disks (Table~\ref{tbl-obspars}) from {\it HST} two-band color maps.

Surface densities were corrected for the effects of inclination, i.e., $\Sigma_{\rm gas}=\Sigma^\prime_{\rm gas}\cos i$, where $\Sigma^\prime_{\rm gas}$ is the observed surface density and $i$ is the estimated disk inclination (derived from kinematic decomposition in $\S$\ref{kin}; see Table~\ref{tbl-kin}). We find that the deprojected gas mass surface densities of our CO-bright galaxies peak between $10^{2.3}$ and $10^{3.2}$ $M_\sun$ pc$^{-2}$, while $\Sigma_{\rm gas}$ averaged over the full disk areas ranges from $\sim10^{1.8}$ to $10^{2.4}$ $M_\sun$ pc$^{-2}$. The observed average surface mass densities agree well with the average $\Sigma_{\rm H\,I+H_2}$ values from the ATLAS$^{\rm 3D}$ survey \citep{ala13,dav13a}. This consistency indicates that radial size and not internal structure is the primary difference between the molecular gas reservoirs for our \mbox{Cycle 2} sample and the ATLAS$^{\rm 3D}$ interferometric sample. We derive representative column densities $N_{\rm H_2}\sim(1-7)\times 10^{22}$ cm$^{-2}$ by averaging the total $M_{\rm H_2}$ values over the elliptical CO-emitting disk areas. If we then assume a disk thickness $h\sim0.1R_{\rm d}$ (e.g., \citealp{bos14}), the corresponding volume-average H$_2$ densities $n_{\rm H_2}$ range between $\sim10$ and 200 cm$^{-3}$. These $N_{\rm H_2}$ values are typical of dusty disks in ETGs, while the estimated $n_{\rm H_2}$ are a couple orders of magnitude lower than predicted from theoretical models \citep{bay13}; the discrepancy can be resolved if the molecular gas is clumpy with a filling factor much below unity. A Galactic $N_{\rm H_2}/A_V=2.21\times10^{21}$ cm$^{-2}$ mag$^{-1}$ \citep{guv09} ratio implies $A_V\sim 5-30$ mag for homogeneous disks (and is highest for the nearly edge-on orientations). As expected for dusty disks that bisect galaxies, these extinction estimates are significantly higher than $A_V$ estimates derived from two-band {\it HST} color maps (with peak $A_V$ ranging between $\sim0.5$ and 1.5 mag for our \mbox{Cycle 2} targets), which method treats the dust disks as foreground (local) obscuration (Figure~\ref{struct}; see also \citealp{kre13}). In $\S$\ref{bhsect}, we discuss large systematic uncertainties that are introduced into the central stellar mass profiles by high extinction across our targets' circumnuclear regions, as well as the implications for BH mass measurements.

\subsubsection{Kinematic Properties}
\label{kin}

\begin{figure*}
\begin{center}
\includegraphics[trim=0mm 0mm 0mm 0mm, clip, width=0.95\textwidth]{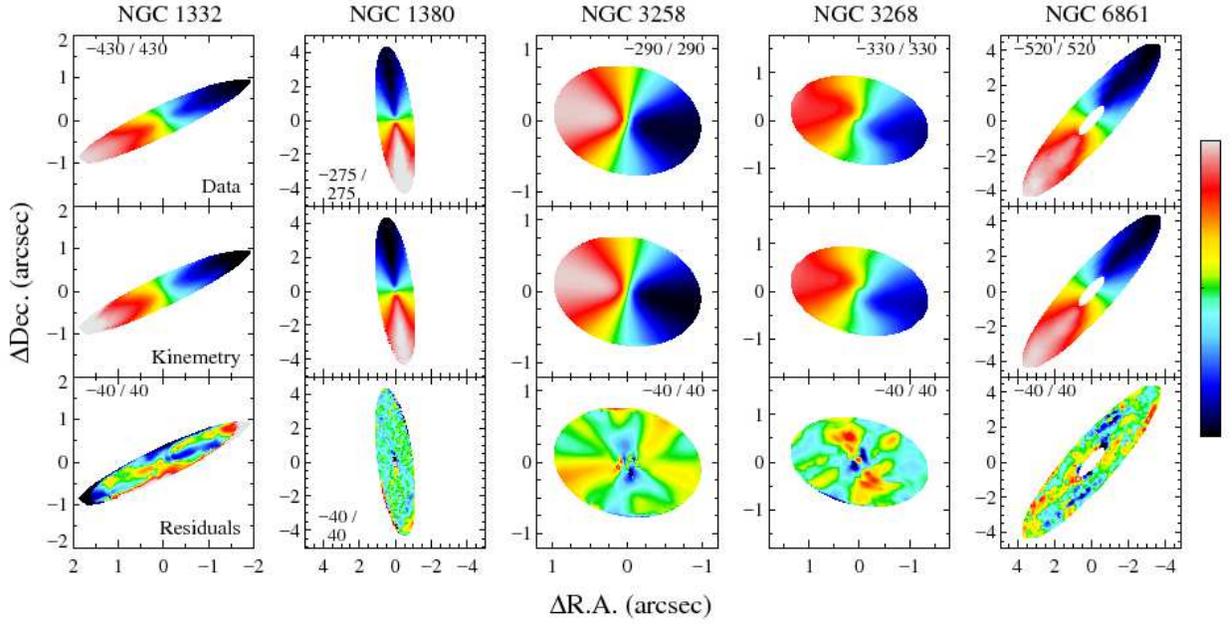}
\caption{Kinemetry modeling results, showing the ({\it top}) highest S/N regions of the $v_{\rm LOS}$ profiles along with ({\it middle}) kinemetry models that only include the circular velocity components and ({\it bottom}) residual maps that indicate generally small ($\lesssim10$ km s$^{-1}$) deviations from circular velocity. The velocity ranges corresponding to the color scaling in the observed $v_{\rm LOS}$ images likewise apply to the kinemetry model maps.}
\label{kin_mod}
\end{center}
\end{figure*}

\begin{figure*}
\begin{center}
\includegraphics[trim=0mm 0mm 0mm 0mm, clip, width=0.95\textwidth]{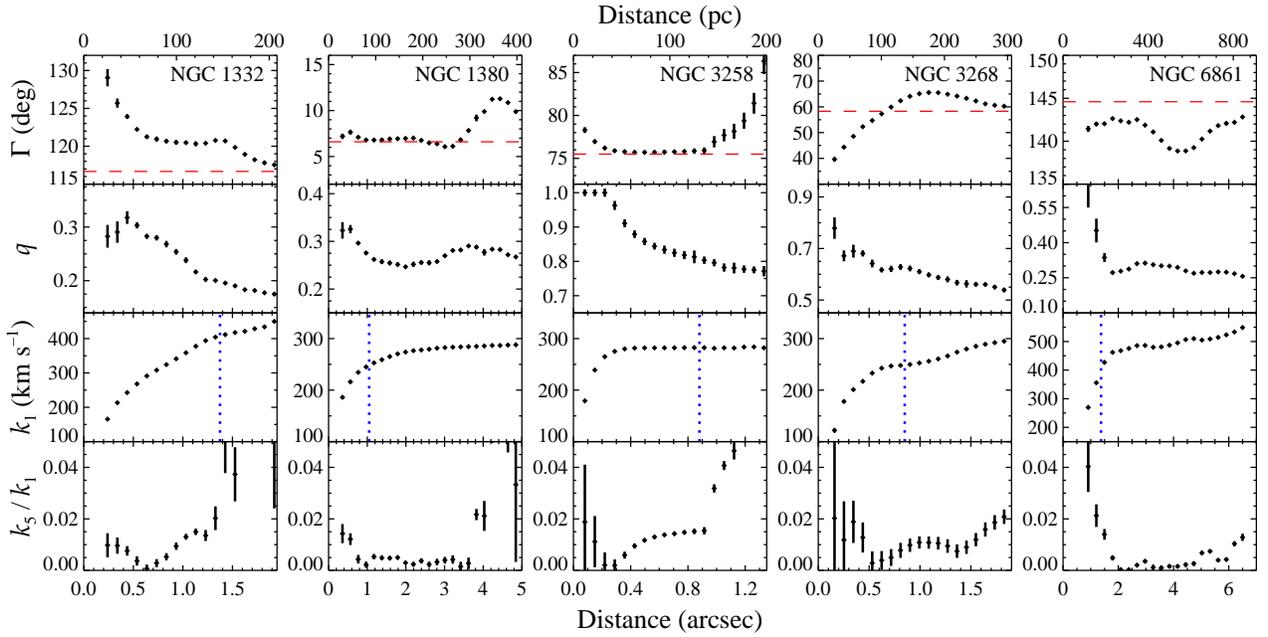}
\caption{Results of the kinemetry expansion of each $v_{\rm LOS}$ field. Kinematic twists of the line of nodes are indicated by a changing position angle $\Gamma$, though the average $\Gamma$ values generally agree very well with the host galaxy stellar photometric major axes position angles (dashed lines). Axis ratio $q$ is related to the gas disk inclination angle, while the $k_5/k_1$ ratio characterizes the level of non-circular motion in the velocity fields. Vertical (dotted) lines alongside the $k_1$ profiles indicate the $2\overline{b}$ ($5\overline{b}$) distance at which we expect $k_1/\sin i\approx v_{\rm c}$ for moderately (highly) inclined disks.}
\label{kin_pars}
\end{center}
\end{figure*}

We characterize the magnitude of kinematic twists using the kinemetry formalism and code of \citet{kra06}. This program performs a harmonic decomposition of a velocity field on elliptical annuli. At each semi-major axis distance $R$ we measure a kinematic position angle $\Gamma$, axis ratio $q$, and harmonic coefficients $k_1$ and $k_5$ for that annulus. Coefficient $k_1$ measures the rotation extremes along the line of nodes (the locus of velocity extrema along elliptical annuli); if $\Gamma (R)$ is roughly constant, then $k_1$ follows $v_{\rm LOS}$ along the semi-major axis. The $k_5$ coefficient quantifies deviations from pure rotation and is useful in conjunction with $k_1$. Kinemetry identifies multiple kinematic elements (e.g., kinematically decoupled components) in the LOS velocity field by the presence of abrupt changes in the PA or flattening ($\Delta \Gamma\gtrsim10\degr$ or $\Delta q\gtrsim0.1$), a double-peaked $k_1$ profile, or a peak in $k_5/k_1$ (at the level of $\gtrsim0.1$). Reconstructed model maps in Figure~\ref{kin_mod} use only the circular velocity terms from kinemetry fits, demonstrating good agreement with the observed velocity fields to (typically) $\lesssim10$ km s$^{-1}$. In Figure~\ref{kin_pars} we show the radial kinemetry fits and in Table~\ref{tbl-kin} we report the best-fitting $v_{\rm sys}$ that minimizes residuals, as well as the average (and total ranges of) $\Gamma$ and $q$ values. The best-fitting recessional velocities are discrepant with those taken from NED by between 20 and 40 km s$^{-1}$. For each of the five disks, the ellipse axis ratio $q$ decreases with increasing $R$. Simple models of disk rotation show a similar decrease in $q$ with radius when incorporating beam smearing. Assuming these disks are thin, we approximate the inclination $i$ by $\cos^{-1}q_{\rm min}$, where $q_{\rm min}$ is the minimum observed axis ratio. These kinematic inclination angles correlate well (to within a few percent) with those estimated from dust disk morphologies (i.e., $\cos^{-1}(b/a)$ from Table~\ref{tbl-obspars}). Kinematic twists tend to be mild ($\lesssim10\degr$) except in the case of \mbox{NGC 3268} -- where $\Delta\Gamma$ approaches $30\degr$ -- and $k_5/k_1$ values are consistently low. Each molecular disk appears to be composed of a single, slightly warped, rotating component.

We use these kinemetry results to estimate the circular speed $v_{\rm c}\approx k_1/\sin i$ and probe the enclosed mass profiles in these five ETGs as a function of radius. At our $\sim50$ pc spatial resolutions, beam smearing mixes emission from far-removed locations in the disk and dilutes the line-of-sight velocities that are intrinsically highest along the line of nodes. For highly-inclined ($i\gtrsim80\degr$) disks, dynamical modeling of the \mbox{NGC 1332} CO(2$-$1) image cube indicates that $k_1/\sin i$ approximates the intrinsic $v_{\rm c}$ to $\lesssim10\%$ accuracy only beyond a radius of $\sim5\overline{b}$ (\citealp{bar16b}; see their Figure 15), where $\overline{b}$ is the geometric mean of the synthesized beam major and minor axis FWHMs. For the more moderately inclined disks ($i\sim45\degr$), modeling the \mbox{NGC 3258} CO(2$-$1) image cube kinematics (Boizelle et al., in prep) demonstrates that $k_1/\sin i\approx v_{\rm c}$ beyond a radius of $\sim2\overline{b}$. We expect that the intrinsic rotation curves of the remaining moderately inclined ($i\approx55\degr$; \mbox{NGC 3268}) galaxy can be approximated at a radius of $\sim 2\overline{b}$, and likewise at a radius of $\sim5\overline{b}$ for the two very inclined ($i\approx75\degr$; \mbox{NGC 1380}, \mbox{NGC 6861}) galaxies we have not yet dynamically modeled. We therefore estimate the masses $M(r<2,5\overline{b})$ of these galaxies enclosed within projected radii of either $2\overline{b}$ or $5\overline{b}$, which span a relatively broad range of $\sim(1.5-8.8)\times10^9$ $M_\sun$ (Table~\ref{tbl-kin}). By design, $\overline{b}\sim r_{\rm g}$ for this \mbox{Cycle 2} sample, so these enclosed masses are dominated by stellar and not BH mass contributions.

We now explore possible disk formation mechanisms in light of these kinemetry results. Dusty, gas-rich disks in ETGs may be either internally generated by {\it in situ} stellar evolution and feedback or externally accreted by gas inflow and merging. In the first scenario, {\it in situ} dusty disk formation may be the result of either the cooling of a hot halo \citep{lag14} or evolved (primarily AGB) stars that inject gas and dust into the interstellar medium at a significant rate (e.g., \citealp{blo95}). Dusty disks that originate from stellar evolution will be closely aligned and co-rotate with the stars. In the second scenario, externally accreted gas streams or mergers with gas-rich galaxies are responsible for the high CO and optically thick dust detection rates. Based on frequent kinematic mismatches ($\sim40\%$ of the ATLAS$^{\rm 3D}$ sample having $\geq30\degr$ discrepancy) between molecular/ionized gas and stars in fainter ($M_K>-24$) ETGs, \citet{dav11b} estimate that at least half of all field ETGs externally acquire their gas. Without ongoing accretion, the gas will relax into the galaxy plane on Gyr timescales \citep{voo15} and become photometrically (though perhaps not kinematically, if the gas rotates counter to the stars; e.g., \citealp{dav11b}) indistinguishable from material entering the gas phase {\it in situ} (following the episodic settling pattern; \citealp{lau05}). However, \citet{lag15} find that smooth gas accretion (e.g., cooling) onto an initially relaxed molecular disk from a misaligned hot halo can promote and maintain large discrepancies between the stellar and gaseous rotation axes. Both scenarios face the complication that dust should be destroyed by thermal sputtering from the hot interstellar medium on relatively short timescales (perhaps up to $100$ Myr; see \citealp{cle10}). Uncertainties in the dust destruction and gas depletion (due to star formation) timescales, as well as in the minor merger rate, make extracting an {\it in situ} formation fraction from surveys of ETGs problematic at best (see \citealp{dav16,bas17}). As \citet{martini13} suggest, these disks may be formed by a combination of internal and external formation mechanisms, such that minor mergers ``seed'' the ETGs with central gas and dust disks, and stellar evolution and halo gas cooling replenishes the cold circumnuclear disks with gas and dust.

\begin{deluxetable}{@{\hskip1pt}c@{\hskip1pt}c@{\hskip1pt}c@{\hskip6pt}c@{\hskip4pt}c@{\hskip1pt}c@{\hskip1pt}c@{\hskip2pt}}[h!]
\tabletypesize{\scriptsize}
\tablecaption{Kinemetry Parameters\label{tbl-kin}}
\tablewidth{0pt}
\tablehead{
\multirow{2}{*}{Galaxy} & \colhead{$v_{\rm sys}$} & \colhead{$\overline{\Gamma}$ / $\Delta\Gamma$} & \multirow{2}{*}{$\overline{q}$ / $\Delta q$} & \colhead{$i$} & \colhead{$\overline{b}$} & \colhead{$M(R<2,5 \overline{b})$} \\
 & \colhead{(km s$^{-1}$)} & \colhead{($\degr$)} &  & \colhead{($\degr$)} & \colhead{(pc)} & \colhead{($10^9$ $M_\sun$)} \\
(1) & (2) & \phantom{+}(3) & (4) & (5) & (6) & (7)
 }
\startdata
NGC 1332 & 1561 & 121.8 / 8.4\phantom{ } & 0.26 / 0.22 & 80 & 29 & 6.20 \\
\rule{0pt}{4ex}NGC 1380 & 1855 & 187.1 / 3.2\phantom{ } & 0.27 / 0.08 & 75 & 17 & 1.54 \\
\rule{0pt}{4ex}NGC 3258 & 2758 & \phantom{ }76.6 / 4.1\phantom{ } & 0.86 / 0.28 & 45 & 65 & 5.03 \\
\rule{0pt}{4ex}NGC 3268 & 2760 & \phantom{ }68.7 / 27.8 & 0.62 / 0.24 & 55 & 69 & 3.14 \\
\rule{0pt}{4ex}NGC 6861 & 2800 & 140.9 / 3.9\phantom{ } & 0.32 / 0.33 & 75 & 36 & 8.83 \\
\enddata
\tablecomments{Molecular gas disk properties of the CO-bright galaxies from kinemetry fits to the CO(2$-$1) $v_{\rm LOS}$ maps. Col. (2): Best-fitting heliocentric systemic velocity which minimized the kinemetry residuals. Col. (3) and (4): The average (and the maximum range in) position angle $\Gamma$ and ellipse axis ratio $q$. Col. (5): Estimated gas disk inclination derived using the thin-disk approximation $i\sim\cos^{-1} q_{\rm min}$ where $q_{\rm min}$ is the minimum axis ratio from the kinemetry modeling. Col. (6): Physical size corresponding to the geometrically-average beam FWHM $\overline{b}$. Col. (7): Estimated mass enclosed within a radius corresponding to 2$\overline{b}$ (or 5$\overline{b}$) for moderately (or highly; $i\gtrsim75\degr$) inclined disks. \\
}
\end{deluxetable}

Our \mbox{Cycle 2} ETGs were selected based on morphologically round dust disks and, for the CO-bright subsample, their molecular gas disks co-rotate with the stars (to within $10\degr$); based on these observations alone, we find that their {\it in situ} formation is as plausible a scenario as the accretion and settling of external gas. For similarly round, clean dust disks (and for dusty ETGs in general), other observational constraints are needed to determine the disk origins. Firstly, if these disks form primarily from the accretion and settling of satellite gas and dust, with little time ($\ll$Gyr) for chemical enrichment from stellar mass loss, their gas-phase metallicities and dust-to-gas ratios should be consistent with those of low-mass galaxies as opposed to the products of late stellar evolution within luminous ETGs. Spatially mapping various metallicity indicators is possible with optical (e.g., \citealp{sto94,pet04}) and mm/sub-mm \citep{bay12} emission line measurements, and a determination of the dust-to-gas ratio, which scales linearly with metallicity, can be made by modeling far-IR (and potentially spatially-resolved, high-frequency ALMA) observations (e.g., \citealp{dav17}). Secondly, deep photometric imaging of galaxies reveals post-merger signatures such as stellar streams or shells (e.g., \citealp{duc15}), and \ion{H}{1} surveys of nearby ETGs often find large-scale disks/rings that are kinematically mismatched with respect to the stellar rotation (e.g., \citealp{ser12}). Even for morphologically round dust disks in ETGs, systems showing recent merger activity on large scales may preferentially host less relaxed circumnuclear disks as demonstrated by the level ($\Delta\Gamma$) of kinematic warping.

\subsubsection{Molecular Gas Stability}
\label{toomre}

Unless stabilized against gravitational fragmentation, a thin, rotating molecular disk is prone to collapse into star-forming clouds. Understanding the stability of the gas disks has implications for both their star formation rates and their disk depletion times \citep{dav14}. \citet{too64} established a local stability criterion $Q$ that can be applied to gas disks based on the gas sound speed $c_s$ and deprojected surface mass density $\Sigma_{\rm gas}$:
\begin{equation}
Q_{\rm gas}\equiv\frac{c_s\kappa}{\pi G \Sigma_{\rm gas}}\;.
\end{equation}
Here, the epicyclic frequency $\kappa$ is defined by $\kappa^2=R\,d\Omega^2/dR+4\Omega^2$, and the angular velocity is $\Omega=v_{\rm c}/R$ with $v_{\rm c}$ the circular velocity at radius $R$. We use $k_1/\sin i$ to approximate $v_{\rm c}$, and reiterate that this approximation is expected to hold past an angular radius of $\sim2\overline{b}$ ($\sim5\overline{b}$) for moderately (highly) inclined disks. Since $c_s$ is much lower than the observed line dispersions (see $\S$\ref{line_profile}), we instead assume that the intrinsic line dispersion contributes more to the gas stability than does thermal pressure. We therefore replace $c_s$ with the minimum observed $\sigma_{\rm LOS}$ for each disk. If the intrinsic line widths of the cold molecular gas increase towards the disk center (for which there is little evidence at high angular resolution; see \citealp{uto15,bar16b}), then $Q_{\rm gas}$ will likewise increase at small radii. Toomre $Q$ values above unity are typically considered stable against gravitational fragmentation, although numerical simulations find that disk instabilities are not fully damped out in regions where $Q$ is only slightly above unity (e.g., \citealp{li05}).

\begin{figure*}
\begin{center}
\includegraphics[trim=0mm 0mm 0mm 0mm, clip, width=0.975\textwidth]{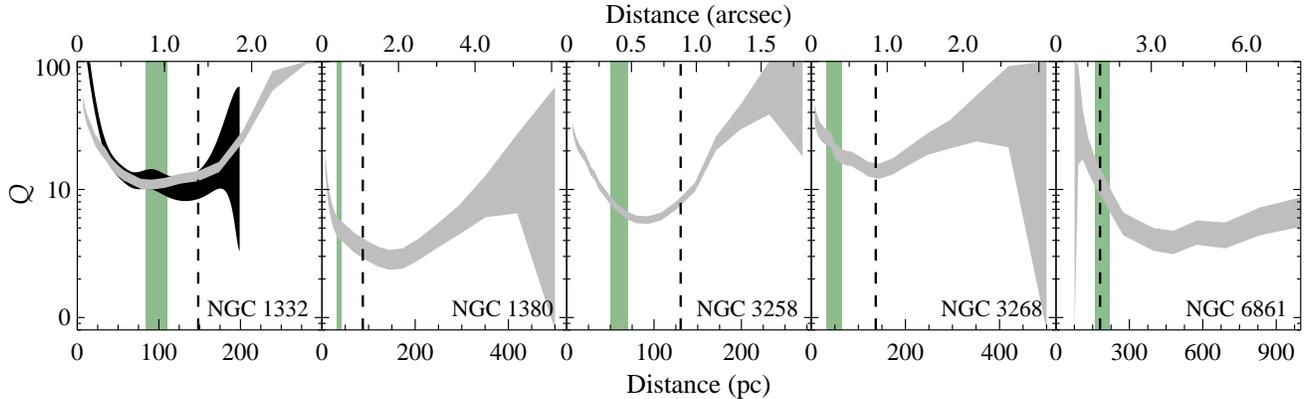}
\caption{Analysis of the molecular gas stability in the CO-bright galaxies using the Toomre $Q$ parameter (shown with uncertainties indicated by the gray shaded regions). These gas disks appear to be formally stable (i.e., $Q>1$) against gravitational fragmentation. The dashed vertical lines correspond to $2\overline{b}$ (or $5\overline{b}$), at which distance from the nucleus the observed LOS velocities are nearly equal to the intrinsic speeds. The shaded vertical regions indicate the projected sphere of influence $r_{\rm g}$, determined by estimating the BH mass using the stellar velocity dispersion in Table~\ref{tbl-obspars} and the $M_{\rm BH}-\sigma_\star$ relationship \citep{kor13}. For \mbox{NGC 1332}, we include a $Q$ profile derived from ALMA \mbox{Cycle 3} $0\farcs044-$resolution gas dynamical modeling (darker regions; \citealp{bar16a}).}
\label{toomreq}
\end{center}
\end{figure*}

For our CO-bright galaxies, we measure $Q_{\rm gas}$ as a function of radius (Figure~\ref{toomreq}) and find that $Q_{\rm gas}>1$ for all radii where $k_1$ is a good proxy for the line-of-sight velocity profile. For all these targets, this formal stability measure persists after we include the uncertainties in $k_1$ and $\Sigma_{\rm gas}$, along with reasonable uncertainties in the inclination angle ($\delta i\sim2\degr$) and minimum line widths ($\delta \sigma_{\rm LOS}\sim1$ km s$^{-1}$). Some portion of the minimum observed $\sigma_{\rm LOS}$ is likely due to rotational broadening and beam smearing ($\S$\ref{line_profile}; see also \citealp{bar16b}), and dynamically modeling the CO-bright disk rotation is under way to determine the intrinsic line widths. Initial modeling results suggests that the observed $\min(\sigma_{\rm LOS})$ exceeds the intrinsic widths by at most a factor of two; as a result, $Q_{\rm gas}$ is expected to decrease slightly but should remain above unity.

Cosmological simulations show that gas disks within ETGs tend to be stable against fragmentation due to larger BH masses and more centrally concentrated stellar profiles than in late-type galaxies (i.e., morphological quenching by bulge growth; \citealp{kaw07,mar09,mar13}). In addition, many ETGs do not possess stellar disks which otherwise would increase the self-gravity of co-spatial gaseous disks. \citet{mar13} suggest that increased shear (often defined by the logarithmic shear rate $\Gamma_{\rm sh}=-d\ln\Omega/d\ln R$; \citealp{jul66}) as a result of bulge growth is the primary stabilizer of ETG gas disks. Since the CO-bright rotation curves are nearly flat from the middle of the disks to the outer edges, the epicyclic frequencies ($\kappa^2=2\Omega^2[2-\Gamma_{\rm sh}]$) are dominated at these radii by high angular speeds and not their logarithmic shear rates. The increase in $Q_{\rm gas}$ near the disk edges is the result of the $\Sigma_{\rm gas}$ profiles declining faster than the epicyclic frequencies. However, a strong intrinsic velocity rise due to the central BH (that is not seen in the $v_{\rm LOS}$ maps due to the $\overline{b}\sim r_{\rm g}$ resolution) will result in much higher $\kappa$ values within the projected $r_{\rm g}$ radius. We explored the inner $Q_{\rm gas}$ profile of \mbox{NGC 1332} using $\sim0\farcs044-$resolution \mbox{Cycle 3} CO(2$-$1) observations that more fully map out the rotation speed within $r_{\rm g}$ \citep{bar16a}. The circular velocity for this profile is derived from the radial mass profile of the best-fitting full-disk dynamical modeling results. At this higher resolution, the observed rotation speed along the major axis is less beam-smeared, and we find that the radial $\kappa$ profile of \mbox{NGC 1332} rises more steeply within the central $\sim50$ pc. This corroborates with previous simulations, showing that the BH within ETGs does further stabilize the central disk regions.

Notwithstanding formal stability ($Q>1$), suppressed star formation still occurs in both simulated and real ETG disks. Based on excess mid-IR flux, \citet{dav14} find evidence for low star formation rates in ETGs with average $\Sigma_{\rm gas}$ and rotational properties similar to that of our sample. We expect the CO emission to be clumpy when observed at high angular resolution, with surface mass densities peaking above the values suggested by the elliptically-averaged $\Sigma_{\rm gas}$ profiles ($\S$\ref{co_sb}; see also \citealp{uto15}). In addition, turbulence and non-circular orbits will create pockets of sufficiently high gas density to be self-gravitating and collapse (see \citealp{hop13}). The presence of a coincident stellar disk lowers the total Toomre $Q$ parameter ($Q_{\rm tot}^{-1}=Q_\star^{-1}+Q_{\rm gas}^{-1}$), although for our S0 galaxies we do not expect the effect to be more pronounced than for the Milky Way (i.e., a factor of $\sim2$ lower in the solar neighborhood; see \citealp{wan94}). Measuring $Q_{\rm tot}$ on sufficiently small scales (i.e., giant molecular cloud sizes) to fully investigate molecular gas stability requires deeper, more highly resolved CO observations to map out the molecular gas distribution and full disk dynamical modeling (with an extinction-corrected stellar mass profile). An alternate (and complementary) method to test disk stability is to directly look for evidence of star forming regions. We identify one such potential location $\sim1\arcsec$ east of the nucleus in \mbox{NGC 3268}. This region is significantly bluer in the {\it HST} color map, while its structure map suggests a light excess. Confirming star formation at this location or elsewhere should be possible using optical IFU data to resolve ionized gas emission, or by modeling sub-arcsecond continuum spectral energy distributions (SEDs).

\subsection{CO-Faint Galaxies}
\label{co-faint}

\begin{figure*}
\begin{center}
\includegraphics[trim=0mm 0mm 0mm 0mm, clip, width=0.95\textwidth]{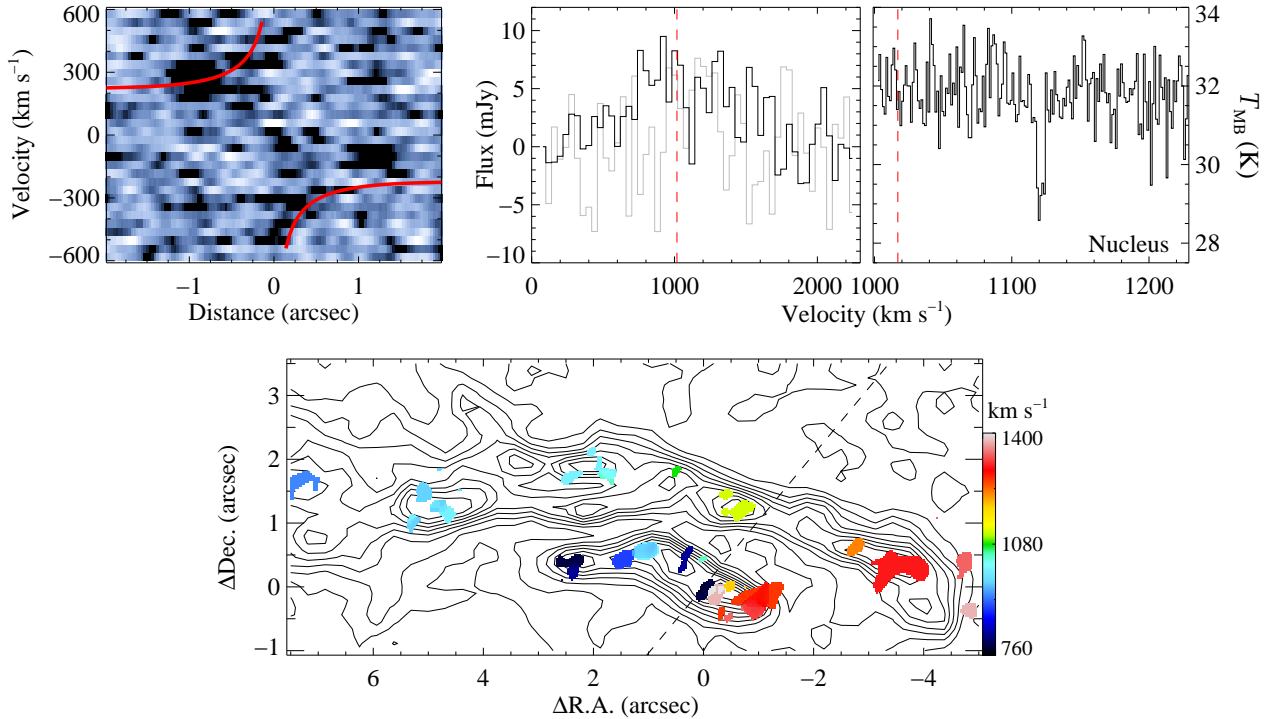}
\caption{\mbox{NGC 4374} CO(2$-$1) emission and absorption characteristics. {\it Top Left}: PVD extracted along the dust disk major axis, using the same negative color scale as the CO-bright target PVDs that displays emission as being darker. For comparison, we include a model $v_{\rm LOS}$ profile (solid curve) assuming the $M_{\rm BH}=10^{8.9}$ $M_\sun$ and a stellar mass profile measured by \cite{wal10}. {\it Top Right}: Velocity profiles formed by ({\it left}) integrating only over the inner, $\sim1\arcsec$ radius dust disk ({\it dark profile}) and only over the outer dust lane ({\it light profile}) in each 40 km s$^{-1}$ channel, and by ({\it right}) measuring the main beam brightness temperature of the nuclear continuum source, imaged in 1.28 km s$^{-1}$ channels. The dashed lines indicate the host galaxy $v_{\rm sys}$ as reported in NED. {\it Bottom}: Contour map of the {\it HST} \mbox{F547M$-$F814W} image (see Figure~\ref{struct}), with coordinates centered on the nuclear 1.3 mm continuum. Regions of the image cube showing faint CO emission are overlaid with a color mapping that corresponds to the channel velocity. The dashed line indicates the larger-scale photometric stellar major axis PA measured from {\it HST} imaging.}
\label{ngc4374_pvd_velchan}
\end{center}
\end{figure*}

\begin{figure*}
\begin{center}
\includegraphics[trim=0mm 0mm 0mm 0mm, clip, width=0.9\textwidth]{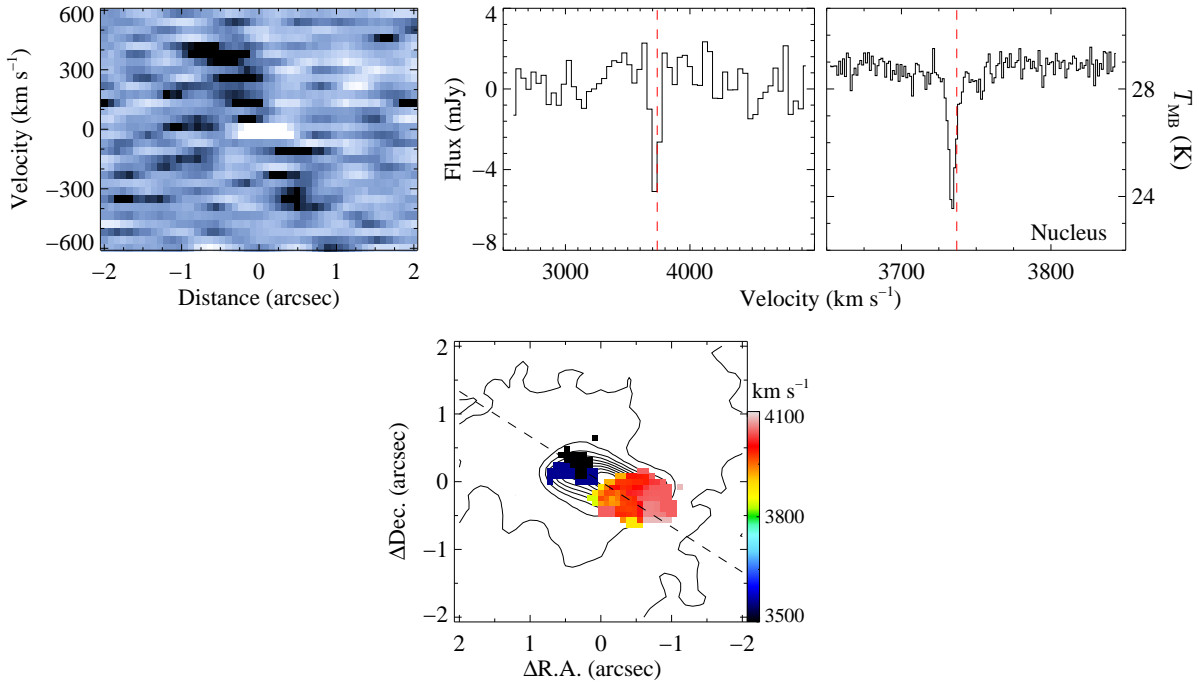}
\caption{\mbox{IC 4296} CO(2$-$1) emission and absorption characteristics. {\it Top Left}: PVD extracted along the dust disk major axis, using the same negative color scale as the CO-bright target PVDs that displays emission as being darker. The bright central feature is the strong nuclear CO absorption. {\it Top Right}: Velocity profiles formed by ({\it left}) integrating only over the inner, $\sim1\arcsec$ radius dust disk in each 40 km s$^{-1}$ channel, and by ({\it right}) measuring the main beam brightness temperature of the nuclear continuum source, imaged in 1.28 km s$^{-1}$ channels. The dashed lines indicate the host galaxy $v_{\rm sys}$ as reported in NED. {\it Bottom}: Contour map of the {\it HST} \mbox{F555W$-$F814W} image (see Figure~\ref{struct}), with coordinates centered on the nuclear 1.3 mm continuum. Regions of the image cube showing faint CO emission are overlaid with a color mapping that corresponds to the channel velocity. The dashed line indicates the larger-scale photometric stellar major axis PA measured from {\it HST} imaging.}
\label{ic4296_pvd_velchan}
\end{center}
\end{figure*}

We constructed velocity profiles for \mbox{NGC 4374} (Figure~\ref{ngc4374_pvd_velchan}) and \mbox{IC 4296} (Figure~\ref{ic4296_pvd_velchan}) by integrating the image cube flux densities in each 40 km s$^{-1}$ channel over their $\sim2\arcsec-$wide optical dust disks (described by the major/minor axes in Table~\ref{tbl-obspars}). The resulting profiles are very different from the symmetric, double-horned shapes that we observe for the CO-bright galaxies. We do detect CO(2$-$1) emission, but at much lower ($\sim10\sigma$ and $\sim5\sigma$) significance than the CO-bright subsample. Their PVDs, extracted along the central dust disk major axes, show faint evidence of disk-like rotation. In both galaxies, careful inspection of the image cubes reveals regions of statistically significant ($\geq3\sigma$ in a 40 km s$^{-1}$ channel) emission across many channels that spatially coincide with the optically thick dust absorption and that show an outline of the full disk rotation. The \mbox{IC 4296 PVD} and approximate velocity maps are consistent with no central high-velocity emission above the background level (although it cannot be ruled out due to the coarse, $0\farcs5$ resolution); in \mbox{NGC 4374}, however, we see hints of a velocity upturn within $r_{\rm g}$ with emission spanning at least 700 km s$^{-1}$ directly across the nucleus. IFU data for \mbox{NGC 4374} show almost no evidence for stellar rotation \citep{ems11}, and the galaxy is expected to be (mildly) triaxial; stellar photometric axes of this slow rotator are misaligned by nearly $90\degr$ with respect to the inner dust disk major axis, the ionized gas kinematics (e.g., \citealp{wal10}), and now seemingly also the cold molecular gas rotation. Long-slit spectroscopy of \mbox{IC 4296} suggests more significant ($\sim100$ km s$^{-1}$) stellar rotation that is roughly consistent with the stellar photometric axes (PA $\sim60-70\degr$; \citealp{efs80}, \citealp{kil86}) and appears to co-rotate to within about $20\degr$ with the faint CO-emission (in Figure~\ref{ic4296_pvd_velchan}).

Even when imaged into 40 km s$^{-1}$ channels, CO(2$-$1) absorption is apparent in \mbox{NGC 4374} and \mbox{IC 4296} against their strong nuclear continuum sources. While extragalactic CO absorption is not frequently observed due to low equivalent widths, ALMA's high-angular resolution and sensitivity allow its detection against some nearby active nuclei (e.g., \citealp{dav14,ran15}). To better characterize central absorption features, we also imaged their visibility data (without continuum subtraction) into cubes with 1.28 km s$^{-1}$ channels. The primary absorption in \mbox{IC 4296} is consistent with the galaxy's assumed systemic velocity (Table~\ref{tbl-obspars}), although there may be a small secondary absorption feature $\sim15$ km s$^{-1}$ redward of $v_{\rm sys}$. The CO(1$-$2) transition seen in the nucleus of \mbox{NGC 4374}, however, is redshifted by $\sim100$ km s$^{-1}$ from the $v_{\rm sys}$ value derived using stellar kinematics \citep{ems04} and by $\sim50$ km s$^{-1}$ with respect to the best-fitting $v_{\rm sys}$ from ionized gas dynamical modeling (e.g., \citealp{wal10}). The absorption lines in these two galaxies have narrow widths ($\lesssim7$ km s$^{-1}$ FWHM) that are only slightly smaller than the minimum line widths in our CO-bright subsample. Maximum absorption depths are $(-3.26\pm0.74)$ and $(-5.25\pm0.35)$ K with integrated absorption intensities $W_{\rm CO(2-1)}$ of $(19.4\pm2.5)$ and $(44.3\pm1.4)$ K km s$^{-1}$ for \mbox{NGC 4374} and \mbox{IC 4296}, respectively. From these $W_{\rm CO(2-1)}$ measurements we estimate corresponding H$_2$ column densities $N_{\rm H_2}=X_{\rm CO}W_{\rm CO(1-0)}$ of $(4.7\pm0.6)\times10^{21}$ cm$^{-2}$ and $(1.1\pm0.3)\times10^{22}$ cm$^{-2}$ after assuming an absorption depth ratio CO(1$-$2)/CO(0$-$1)$\approx0.6$ (based on $^{13}$CO absorption line ratios; e.g., see \citealp{eck90}) and a CO-to-H$_2$ conversion factor $X_{\rm CO}$ corresponding to $\alpha_{\rm CO}=3.1$ $M_\sun$ pc$^{-2}$ (K km s$^{-1}$)$^{-1}$ (see $\S$\ref{co_sb}; \citealp{san13}). These $N_{\rm H_2}$ estimates are much more uncertain than indicated by the formal associated uncertainties, since the $\alpha_{\rm CO}$ and CO(1$-$2)/CO(1$-$0) values carry large systematic uncertainties, and faint nuclear CO emission may somewhat dilute these equivalent widths.

While the line profiles are dominated by noise on a pixel-by-pixel basis, we estimate the total CO(2$-$1) emission flux by integrating the \mbox{NGC 4374} and \mbox{IC 4296} velocity profiles outside of the absorption features and over $\pm520$ and $\pm480$ km s$^{-1}$ from $v_{\rm sys}$, respectively, over which faint emission is detected in the PVDs. The central disks of these CO-faint galaxies have low integrated CO(2$-$1) fluxes ($<5$ Jy km s$^{-1}$; see Table~\ref{tbl-linemeas}) with corresponding total (molecular and helium) gas masses of $\sim6\times10^6$ $M_\sun$ (refer to $\S$\ref{co_sb}). For completeness, we also measure a velocity profile for the larger-scale, outer dust lane in \mbox{NGC 4374} (included in Figure~\ref{ngc4374_pvd_velchan}) and find a possible CO detection with $I_{\rm CO(2-1)}=(1.6\pm1.1)$ Jy km s$^{-1}$ when integrated over the same $\pm520$ km s$^{-1}$ velocity range. Previous single-dish (CSO: \citealp{kna96}; IRAM 30m: \citealp{com07}, \citealp{oca10}) observations of \mbox{NGC 4374} found at best only a tentative CO detection (corresponding to $I_{\rm CO(2-1)}=8.8\pm1.6$ Jy km s$^{-1}$), while similar (APEX: \citealp{pra10}) observations of \mbox{IC 4296} found only an upper limit of $I_{\rm CO(2-1)}<20$ Jy km s$^{-1}$ (corresponding to $M_{\rm H_2}\lesssim10^{8.7}$ $M_\sun$). {\it IRAS} far-IR data, which are sensitive to thermal emission from both clumpy, optically opaque dust as well as additional diffuse dust, suggest total dust masses of $\sim10^{4.6-5.0}$ $M_\sun$ in these two galaxies \citep{oca10,amb14}. Using a fiducial 1:100 dust-to-gas mass ratio \citep{rui02}, our $M_{\rm H_2+He}$ measurements for these $\sim2\arcsec-$wide dusty disks (and even when including the \mbox{NGC 4374} outer dust lane) are consistent with $\sim10^7$ $M_\sun$ gas mass estimates from thermal dust emission.

When these $M_{\rm H_2}$ masses are averaged over the $\sim1\arcsec$ radius optical dust disk regions, the estimated H$_2$ column densities are $\sim4.0\times10^{21}$ cm$^{-2}$ for \mbox{NGC 4374} and $\sim9.8\times10^{21}$ cm$^{-2}$ for \mbox{IC 4296}. Without even considering the large total flux uncertainties ($\sim15-25\%$), these CO emission-derived $N_{\rm H_2}$ estimates are consistent with those from integrated absorption line intensities. After correcting for inclination effects (assuming $i\sim75\degr$), the deprojected H$_2+$He surface mass densities are fairly low ($\sim10^{1.7}$ and $\sim10^{1.4}$ $M_\sun$ pc$^{-2}$, respectively) compared to the CO-bright subsample (whose average $\Sigma_{\rm gas}$ values range from $10^{1.8}$ to $10^{2.4}$ $M_\sun$ pc$^{-2}$; see $\S$\ref{co_sb}).

It is not yet clear why the central disks of these two galaxies are underluminous in CO(2$-$1) compared to the CO-bright subsample. Intriguingly, the dusty disks in \mbox{NGC 4374} and \mbox{IC 4296} color maps show redder $V-I$ colors than the similarly sized and inclined \mbox{NGC 1332} dust disk, suggesting these CO-faint disks may possess larger dust column densities. Using their average $N_{\rm H_2}$ values, \mbox{NGC 4374} and \mbox{IC 4296} have $A_V\leq4$ mag extinction estimates. However, the dusty disk in \mbox{NGC 1332} has a very large $A_V\sim30$ mag estimated from its average $N_{\rm H_2}$ and possesses over an order of magnitude larger H$_2$ reservoir. Both \mbox{NGC 4374} \citep{bow97} and \mbox{IC 4296} \citep{ann10} possess significant nuclear ionized gas emission, which may indicate that their dust disks are associated with large \ion{H}{1} or ionized gas reservoirs. Low $\Sigma_{\rm H_2+He}$ values suggests that the $\Sigma_{\rm H\,I}$ and $M_{\rm H\,I}$ values may be on parity with their molecular gas counterparts. These two Fanaroff-Riley (FR) type-I radio galaxies \citep{fan74} have both large \citep{kil86,lai87} and small-scale \citep{har02,pel03} radio jets while our CO-bright galaxies are not known to contain radio jets \citep{lar81}. \mbox{NGC 4374} and \mbox{IC 4296} show strong 1.3 mm nuclear continua relative to the CO-bright targets (see Table~\ref{tbl-contmeas}), and their extended 1.3 mm continuum profiles may be small-scale ($\sim100$ pc; see $\S$\ref{contsb}) nuclear jets. The luminosities and energetics of active nuclei may dissociate a significant portion of the H$_2$ \citep{mul13}, giving rise to smaller molecular gas reservoirs, or excite the CO into higher$-J$ transitions more efficiently than in normal galaxies (compare to the \mbox{NGC 4151} active nucleus, which has significant warm H$_2$ rovibrational emission yet is undetected in low$-J$ CO transitions; \citealp{sto09,dum10}).

From optical {\it HST} imaging, nuclear dust features are found in a third to half of nearby FR sources in the 3CR and B2 radio catalogs, frequently appearing as morphologically regular disks \citep{dek00,cap00,rui02}. CO-detected FR galaxies have an average $M_{\rm H_2}$ of $\gtrsim10^8$ $M_\sun$ (although with only $\sim50\%$ detection rates; e.g., \citealp{oca10}). Our two CO-faint galaxies are either at or (in the case of \mbox{IC 4296}) an order of magnitude below the CO detection threshold in these earlier surveys. We suspect that a large fraction of previously CO-undetected FR galaxies will show emission (with at least partially resolved gas kinematics) when observed with ALMA in relatively short ($\sim1$ hr per target) on-source integrations. Such observations will allow investigation of the physical conditions (e.g., molecular gas kinematics, disk dynamical coldness) of the large-scale dusty disks that power these radio galaxy nuclei. Furthermore, ALMA CO observations of a statistically unbiased sample of radio-loud and radio-quiet ETGs will reveal whether there exists any correlation between the molecular gas mass or surface density and the nuclear radio (and mm/sub-mm) continuum intensity.

\section{Continuum Properties}
\label{contprop}

All seven \mbox{Cycle 2} targets show significant continuum emission concentrated near the galaxy centers. For our CO-bright (CO-faint) targets, the locations of the continuum peaks listed in Table~\ref{tbl-contmeas} agree with the line kinematic centers to $\lesssim \overline{b}/5$ ($\sim\overline{b}/2$). Since at least some level of nuclear activity is seen in most ETGs \citep{ho97,nyl16}, the centrally peaked continua in our sample may be dominated by emission from low-luminosity AGN (LLAGN). Other sources may include thermal and free-free emission from (AGN or stellar) photoionized regions, and thermal emission from cool dust grains. Through a combination of {\it uv} plane modeling and spectral fitting, we investigate the continuum emission properties of our \mbox{Cycle 2} sample.

\begin{deluxetable*}{lccccccc}[h!]
\tabletypesize{\scriptsize}
\tablecaption{Continuum Properties\label{tbl-contmeas}}
\tablewidth{0pt}
\tablehead{
\multirow{2}{*}{\phantom{m}Galaxy} & \multicolumn{2}{c}{Peak Position (J2000)} & \colhead{RMS Noise} & \colhead{$S_{\rm nuc}$} & \multirow{2}{*}{$\alpha_{\rm nuc}$} & \colhead{$S_{\rm ext}$} & \multirow{2}{*}{$\alpha_{\rm ext}$}\\
 & \colhead{$\alpha$} & \colhead{$\delta$} & \colhead{($\mu$Jy beam$^{-1}$)} & \colhead{(mJy)} & & \colhead{(mJy)} &  \\
\multicolumn{1}{c}{(1)} & (2) & (3) & (4) & (5) & (6) & (7) & (8)
 }
\startdata
NGC 1332 & 03:26:17.234 & $-$21:20:06.81 & 28.2 & \phantom{ }8.53 $\left(^{+0.88}_{-1.13}\right)$ & $-$0.44 (0.11) & 2.05 $\left(^{+0.50}_{-0.35}\right)$ & \phantom{$-$}2.60 (2.05) \\
\rule{0pt}{4ex}NGC 1380 & 03:36:27.573 & $-$34:58:33.84 & 26.8 & \phantom{ }6.05 $\left(^{+0.64}_{-1.14}\right)$ & \phantom{$-$}1.62 (0.17) & 2.35 $\left(^{+0.67}_{-0.40}\right)$ & \phantom{$-$}6.10 (1.95) \\
\rule{0pt}{4ex}NGC 3258 & 10:28:53.550 & $-$35:36:19.78 & 21.4 & \phantom{ }0.44 $\left(^{+0.18}_{-0.24}\right)$ & \phantom{$-$}0.90 (2.09) & 0.47 $\left(^{+0.27}_{-0.09}\right)$ & \phantom{$-$}3.18 (2.69) \\
\rule{0pt}{4ex}NGC 3268 & 10:30:00.654 & $-$35:19:31.55 & 27.0 & \phantom{ }3.45 (0.61) & $-$0.28 (0.30) & $<0.39$ & $-$ \\
\rule{0pt}{4ex}NGC 4374 & 12:25:03.745 & \phantom{$-$}12:53:13.11 & 53.8 & 126.0 $\left(^{+13.1}_{-16.1}\right)$ & $-$0.21 (0.02) & 3.47 $\left(^{+7.17}_{-0.68}\right)$ & $-$0.56 (1.03) \\
\rule{0pt}{4ex}NGC 6861 & 20:07:19.469 & $-$48:22:12.47 & 34.1 & \phantom{ }22.6 (2.29) & $-$0.27 (0.05) & $<2.04$ & $-$ \\
\rule{0pt}{4ex}IC 4296 & 13:36:39.040 & $-$33:57:57.18 & 164.4 & 212.1 $\left(^{+21.4}_{-22.0}\right)$ & \phantom{$-$}0.10 (0.02) & 2.59 $\left(^{+1.28}_{-0.69}\right)$ & $-$ \\
\enddata
\tablecomments{Spatial and spectral properties of the nuclear and extended continua. Cols. (2) and (3): Nuclear continuum centroid location, in hh:mm:ss and dd:mm:ss format, from imaging prior to any any self-calibration. The astrometric precision of these observations is $\sim0\farcs01$. Col. (4): The rms background noise in the naturally-weighted continuum images. Col. (5): Peak flux densities at an observed frequency of $\sim236$ GHz. Asymmetric uncertainties for $S_{\rm nuc}$ and $S_{\rm ext}$ arise from first fitting the continua as completely unresolved sources in the $uv$ plane and afterwards requiring that the extended emission monotonically decreases with radius. Col. (6): Spectral indices (where $S_\nu\propto \nu^{\alpha}$) of the unresolved emission using measurements of the peak continuum levels in separate spectral windows $\gtrsim6$ GHz apart. Col. (7): Flux densities (or $3\sigma$ upper limits) of the extended continuum  profiles, integrated over the optical dust disk regions. The uncertainties in $S_{\rm nuc}$ and $S_{\rm ext}$ tend to be dominated by the ($\sim10$\%) uncertainty in the absolute flux calibration. (8): Spectral indices of the extended emission using $S_{\rm ext}$ measurements of the extended continuum levels in separate spectral windows.}
\end{deluxetable*}

\subsection{Continuum Surface Brightness Modeling}
\label{contsb}

\begin{figure}
\begin{center}
\includegraphics[trim=0mm 0mm 0mm 0mm, clip, height=0.8\textheight]{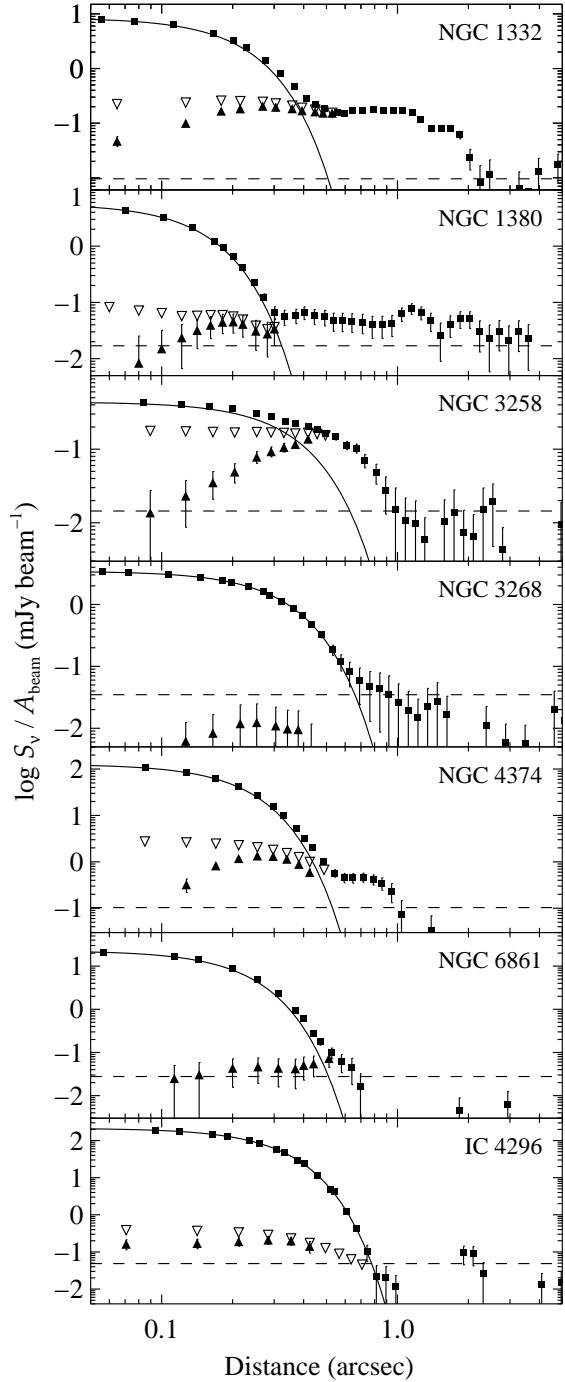}
\caption{Continuum surface brightness profiles of the naturally weighted continuum images. The central peak component is modeled as an unresolved source in the $uv$ plane, as represented by the accompanying scaled synthesized beam profiles (solid curves). The extended continua are first measured (filled diamonds) after removing the best-fit point source from the visibilities. When these fits yield significant nuclear residuals (in all cases except for \mbox{NGC 3268} and \mbox{NGC 6861}), we subtract slightly lower, but still centrally-dominant, point source flux densities in the $uv$ plane until the image plane residuals are a monotonically decreasing function of radius (open diamonds). Horizontal (dashed) lines show the rms sidelobe noise levels which are determined by measuring the surface brightness beyond the optical dust disk radii.}
\label{cont_sb}
\end{center}
\end{figure}

When imaged with natural weighting, most of our targets (with the exception of \mbox{NGC 3268} and \mbox{NGC 6861}) show significant extended emission (see Figure~\ref{struct}). We model and subtract the central point sources in the $uv$ plane prior to re-imaging the visibilities and isolating the extended continua. This point-source subtraction may leave a slight depression in the center of the extended continuum emission (e.g., for \mbox{NGC 3258}) and can also introduce imaging artifacts (in the case of \mbox{IC 4296}). For the targets with significant extended emission, we account for possible over-subtraction of the point source by performing the {\it uv}-plane point-source subtraction with consecutively lower unresolved flux density values until the extended emission in the image plane has a roughly monotonically decreasing central profile (Figure~\ref{cont_sb}). We take this as an appropriate upper bound to the extended continuum emission.

We determine the central unresolved ($S_{\rm nuc}$) and extended ($S_{\rm ext}$) continuum emission values (Table~\ref{tbl-contmeas}) by measuring the peak intensities from {\it uv}-plane modeling and by integrating the extended emission, respectively. For \mbox{NGC 3268} and \mbox{NGC 6861}, we integrate over their optical dust disk regions to determine $3\sigma$ upper limits on their extended emission. For the remainder of the sample, we only integrate the extended emission out to the radius where negative sidelobes begin to decrease the $S_{\rm ext}$ values. Uncertainties in the tabulated flux densities include changes in $S_{\rm nuc}$ and $S_{\rm ext}$ when we force the point-source subtraction to produce centrally peaked extended continuum profiles. We separate these $S_{\rm nuc}$ measurements into mm-faint and mm-bright ($S_{\rm nuc}>100$ mJy) nuclear sources for convenience during later discussion. We note that the two mm-bright galaxies \mbox{NGC 4374} and \mbox{IC 4296} are also the brightest radio (FR I) sources.

The resolved continuum profiles of \mbox{NGC 1332}, \mbox{NGC 1380}, and \mbox{NGC 3258} follow the shapes and orientations of their optical dust disks. Since continuum emission above $\sim200$ GHz is usually dominated by thermal reprocessing of starlight (e.g., \citealp{con92}), this consistency suggests a thermal origin for $S_{\rm ext}$. Dust absorption is apparent in {\it HST} images, so the dominant thermal source is likely cold ($T\sim30$ K; \citealp{ben03,gal12}) dust grains. However, LLAGN and central star formation photoionize and heat their surroundings, and Bremsstrahlung emission from ionized atomic gas with electron temperatures of $\sim10^4$ K may also contribute to the mm/sub-mm continuum emission. Long-slit spectroscopy of \mbox{NGC 1380}, \mbox{NGC 3258}, \mbox{NGC 3268}, \mbox{NGC 4374}, and \mbox{IC 4296} reveals photoionized gas with low H$\beta$ line luminosities ($\lesssim10^{38}$ erg s$^{-1}$) and low densities ($n_e\lesssim10^3$ cm$^{-3}$) within the central $2\arcsec\times2\arcsec$ \citep{ann10} that are typical of ``dwarf'' Seyfert nuclei \citep{ho97}. Based on Balmer flux decrements, the extinction in the observed ionized gas regions is low ($A_V\lesssim1.5$ mag; \citealp{ann10}), suggesting that the ionized gas emission is physically distinct from the molecular gas regions. In the scenario where the ionized gas exists around and between clumpy, optically thick ($A_V \gtrsim10$) clouds, at most half of the intrinsic H$\beta$ line emission will be heavily obscured (in the case of a uniformly opaque molecular disk). We therefore expect that the observed $L({\rm H}\beta)$ values underestimate the total H$\beta$ luminosity by no more than a factor of two. \citet{ulv81} derive a relationship between the Balmer narrow line flux and the low-frequency thermal flux density of low density environments. For a $\sim10^4$ K emitting region, we expect that the nuclear photoionized gas in these galaxies should produce at most only a few $\mu$Jy of thermal continuum emission at 230 GHz. Thermal emission from dust and not photoionized gas is therefore the most likely $S_{\rm ext}$ source for the CO-bright galaxies \mbox{NGC 1332}, \mbox{NCG 1380}, and \mbox{NGC 3258}.

The continuum residuals of the mm-bright sources \mbox{NGC 4374} and \mbox{IC 4296} extend roughly perpendicular to the orientation of their optical dust disks, and are therefore not likely related to thermal emission. Instead, these extended features may trace resolved synchrotron jets on tens of pc scales. Sub-arcsecond resolution radio observations show an inner jet position angle of $\sim170-180\degr$ for \mbox{NGC 4374} \citep{jon81,har02} and $\sim140\degr$ for \mbox{IC 4296} \citep{pel03}, while we estimate the orientations of the residual continuum emission to be at about $135\degr$ and $140\degr$, respectively. Continuum observations at additional (esp. lower) frequencies of equivalent or greater angular resolution should allow for a more confident interpretation of the mm-bright $S_{\rm ext}$ profiles.

\subsection{Continuum Spectral Fitting}

The continua within the central beam areas are heavily dominated by unresolved emission for all targets except \mbox{NGC 3258} (although its peak unresolved emission still exceeds the resolved continuum by at least a factor of two). To measure crude mm-wavelength spectral indices, we imaged the continuum-only spectral windows into individual continuum maps with frequency separations of $\sim20$ GHz. The unresolved nuclei were again subtracted in the {\it uv} plane (as described in $\S$\ref{contsb}) to determine $S_{\rm nuc}$ and $S_{\rm ext}$ values for each spectral window image. Nuclear and extended spectral indices were measured at an observed $\sim236$ GHz by modeling the unresolved and (where detected) resolved continuum emission as a power law ($S_\nu\propto\nu^\alpha$).

The unresolved 1.3 mm continuum spectral indices $\alpha_{\rm nuc}$ of our \mbox{Cycle 2} sample range from $-0.44$ to 1.62 (see Table~\ref{tbl-contmeas}), and are roughly consistent with both the radio spectral indices and flux densities of other LLAGNs \citep{nag01,nem14}. These spectral slopes are likely to result from some nuclear thermal dust emission (with index $\alpha_{\rm th}\sim3-4$; see \citealp{dra03,dra06}) surrounding a compact active nucleus, consisting of an accretion disk and perhaps a synchrotron jet (with $\alpha_{\rm syn}\sim-0.7$). In starburst galaxies (e.g., M82; \citealp{car91}, \citealp{con92}), free-free emission from ionized stellar winds may also contribute (with $\alpha_{\rm ff}\sim-0.1$ to 0.4 at mm/sub-mm wavelengths; see \citealp{pas12}), but our \mbox{Cycle 2} galaxies do not show any signs of significant nuclear star formation. The central engines of LLAGN are expected to be powered by hot accretion flows (such as advection-dominated accretion; \citealp{nar94}, \citealp{nar95}). Given plausible accretion model parameters and the previous $M_{\rm BH}$ measurements for this sample, our measured $\alpha_{\rm nuc}$ values are consistent with the predicted 1.3 mm spectral index for hot accretion flows (\citealp{yua14}; see their Figure 1); however, a mixture of optically thin synchrotron and thermal emission can equally well reproduce this range in spectral slopes. Additional ALMA ($\sim100-900$ GHz) continuum measurements, along with sub-arcsecond radio flux densities, would enable us to confidently identify the dominant nuclear mm/sub-mm emission sources (e.g., see \citealp{ala15}).

Based on the spatial correspondance between the CO emission or radio jet orientation, the extended continuum profiles in our sample appear to originate from either thermal (with $\alpha_{\rm th}\sim3-4$) or nonthermal ($\alpha_{\rm syn}\sim-0.7$) sources, respectively. The resolved continuum spectral indices $\alpha_{\rm ext}$ for \mbox{NGC 1332}, \mbox{NGC 1380}, and \mbox{NGC 3258} (in Table~\ref{tbl-contmeas}) are all consistent with a purely thermal origin. As the cold dust in these disks is likely at a temperature of $\sim30$ K, the 1.3 mm flux densities lie very far down on the Rayleigh-Jeans tail of thermal emission. Measuring $S_{\rm ext}$ across higher frequency ALMA bands will allow for more accurate $\alpha_{\rm ext}$ values and provide more useful constraints on the physical state of the dust. The extended continuum of \mbox{NGC 4374} is inconsistent with a predominantly thermal source; due to large uncertainties inherent in extracting a weak extended source that is barely resolved, we cannot state with certainty that its $\alpha_{\rm ext}$ measurement is strong evidence for a resolved synchrotron jet.

\section{Prospects for BH Mass Measurements}
\label{bhsect}

The primary motivation for these \mbox{Cycle 2} observations is to detect high-velocity rotation from within $r_{\rm g}$. In cases where the central velocity upturns are unambiguously detected, we plan to propose for higher spatial resolution ALMA observations to fully map and dynamically model the molecular gas kinematics. \citet{bar16b,bar16a} have demonstrated the efficacy of this two-stage process for \mbox{NGC 1332}, measuring its $M_{\rm BH}$ to 10\% precision. In comparison, BH mass measurements via stellar or ionized gas dynamical modeling typically produce 20-50\% uncertainties on $M_{\rm BH}$ \citep{kor13}. High-resolution ALMA observations can provide crucial cross-checks for BH masses measured by stellar dynamical modeling or other methods: for example, \citet{bar16a} use higher-resolution ALMA data to rule out a previous stellar dynamical BH mass measurement for \mbox{NGC 1332} at high confidence. While ALMA has the capacity to highly resolve $r_{\rm g}$ and produce tight BH mass constraints for nearby galaxies, the majority of ETGs are not suitable targets for such studies. Roughly half of all ETGs show optically thick dust features, while only about 10\% of these have morphologically regular dust disks (e.g., \citealp{van95,tra01}). Our \mbox{Cycle 2} findings suggest that the majority of such dust disk candidates should be detectable in CO with relatively short integration times using ALMA. To enable BH mass measurements, the observed disks need to be in circular rotation with low turbulence (low intrinsic $\sigma/v_{\rm c}$) and have high-velocity emission originating from within $r_{\rm g}$. Only 3/5 of our CO-bright targets show central velocity upturns on spatial scales comparable to $r_{\rm g}$, with the possible addition of the low S/N (but potentially large) velocity upturn in one CO-faint galaxy (\mbox{NGC 4374}). In addition, highly inclined disks are more difficult to model due to strong beam smearing effects that dilute the major axis rotation when the sphere of influence projected along the minor axis ($r_{\rm g}\cos i$; see \citealp{bar16b}) is not well resolved. Based on these results, we expect that only a few percent of nearby ETGs will be good targets for precision BH mass measurements with ALMA. The barrier here is slightly higher than finding regular kinematics in circumnuclear ionized gas disks for BH mass measurements (though for different reasons), where the success rate without any pre-selection is somewhat less than 20\% \citep{kor13}. When rapid molecular rotation is detected within $r_{\rm g}$, however, the low turbulence and modest levels of disk warping make these ETG disks prime candidates for high-resolution ALMA observations.

In the sample PVDs we do not detect any CO(2$-$1) emission with deprojected velocity over $\sim600$ km s$^{-1}$; this suggests that significant CO emission does not originate from deep within the BH sphere of influence. Observing denser gas tracers (including higher-$J$ CO lines) at similar (or higher) angular resolutions will clarify whether molecular gas is abundant in the central few tens of parsecs of CO-bright ETGs. If such tracers are absent, the mm continuum sources we detect in all of our \mbox{Cycle 2} sample are likely candidates to have dissociated or disrupted the inner molecular disks. Observations of dusty disks in additional ETGs may reveal whether a correlation exists between the nuclear continuum intensity $S_{\rm nuc}$ and the presence of holes (as with \mbox{NGC 6861}, which has the brightest $S_{\rm nuc}$ of the CO-bright subsample) or potential depressions (as with \mbox{NGC 1332}) in the inner emission-line surface brightness profiles.

For late-type galaxies, which tend to be gas-rich, a similar fraction may also be good candidates for BH mass measurement, and cold molecular gas tracers have already been observed at arcsecond to sub-arcsecond scales in some nuclei (e.g., \citealp{gar14,oni15,gar16}). While observations of late-type galaxies are now approaching the resolution required to unambiguously detect the BH kinematic influence, complications such as spiral structure and radial flows may cause strong deviations from purely circular rotation that will be difficult to model (e.g., \citealp{gar16}). Late-type galaxies also tend to have much lower $M_{\rm BH}$ values than do luminous ETGs, and the smaller BH spheres of influence for nearby spirals are more difficult to resolve with ALMA.

Especially for ALMA observations which do not highly resolve molecular rotation within $r_{\rm g}$, accurate stellar profiles are essential when modeling gas disk dynamics. {\it HST} imaging of these ETGs indicates that dust obscuration extends from roughly the outer edge of CO emission down to the nuclei. If the gas and dust are uniformly distributed in these disks, the average H$_2$ column densities derived from CO intensities suggest average visual extinctions of $A_V\sim5-30$ mag; using standard extinction transformations \citep{rie85}, we expect between 0.5 and 3 mag of $K-$band extinction of the stellar light that originates from {\it behind} the dust disks of our \mbox{Cycle 2} targets. The slopes of the central stellar luminosity profiles will therefore remain somewhat uncertain even if high spatial resolution NIR imaging is obtained. When modeling our \mbox{Cycle 2} data, we find that uncertainties in the stellar slope tend to produce changes in the best-fitting $M_{\rm BH}$ that are larger than the formal BH mass uncertainties (\citealp{bar16b}; Boizelle et al., in prep.). To minimize systematic errors from an uncertain stellar mass profile, the optimal situation is to obtain ALMA observations that highly resolve the BH sphere of influence along both the disk major and minor axes. Ideally, this requires several independent resolution elements across the projected minor axis $r_{\rm g}\cos i$. Given the need for reasonable S/N while highly resolving $r_{\rm g}$, accurately measuring $M_{\rm BH}$ in ETGs is nontrivial, even with ALMA. Even without highly resolved observations, or in disks that show no central, rapid rotation, dynamical modeling is still capable of constraining or placing upper limits on $M_{\rm BH}$ after accounting for reasonable uncertainties in the inner stellar slope.

Gas dynamical modeling of ALMA observations has the potential to produce some of the most accurately measured BH masses \citep{bar16a} after the Milky Way BH \citep{ghe08,gil09} and a few determined using H$_2$O maser disks (e.g., \citealp{miy95,kuo11}). Since only a few percent of nearby ETGs are likely to be good candidates for these high-resolution studies, careful sample selection is required. We have demonstrated that target pre-selection based on regular dust morphology produces a very high CO detection rate. ALMA observations with spatial resolution roughly matching $\sim r_{\rm g}$ are sufficient to identify central, fast-rotating gas and justify follow-up, high-resolution observations. This two-stage process is necessary to identify high-quality targets for long-baseline studies and to make optimal use of precious ALMA observing time.

\section{Summary and Conclusions}

We have presented ALMA Band 6, $\sim0\farcs3-$resolution observations of seven ETG nuclei that were selected based on their optical dust disk morphologies. Five of the seven are detected in CO(2$-$1) at high S/N, arising from disks with total gas masses of $\sim10^8$ $M_\sun$ and average deprojected mass surface densities on the order of 100 $M_\sun$ pc$^{-2}$. The remaining two show faint but unambiguous CO emission with much lower ($\sim10^6$ $M_\sun$ and $\lesssim50$ $M_\sun$ pc$^{-2}$) total gas masses and surface mass densities. These CO-faint galaxies also show strong CO(2$-$1) absorption coincident with their mm-bright nuclear continua. In all cases, line-of-sight velocity maps indicate nearly circular rotation about the disk centers and confirm that disky dust morphology is a good predictor of regular molecular gas rotation. Four of the seven disks show velocity upturns in CO emission within $\sim r_{\rm g}$ that indicate partial resolution of the BH kinematical signature. Warping of the gas disks, although present in every CO-bright case, is both smooth and generally of low magnitude ($\lesssim10\degr$, except for \mbox{NGC 3268}). Based on low line dispersions and regular gas kinematics, we find that our CO-bright targets possess dynamically cold gas disks. We explore the stability of these molecular disks against gravitational fragmentation using the Toomre $Q$ test. For the outer regions of these disks, $Q_{\rm gas}>1$; based on higher-resolution observations of \mbox{NGC 1332} we expect this stability measure to extend within $r_{\rm g}$. We anticipate that star formation will be suppressed (although not necessarily absent) in these systems, and the corresponding star formation surface densities for these systems will lie below the standard Kennicutt-Schmidt relationship of late-type galaxies (as is found for the ATLAS$^{\rm 3D}$ ETG sample; \citealp{dav14}).

Each galaxy in our sample possesses a centrally-dominant, unresolved nuclear 1.3 mm continuum source that is most likely accretion powered. The $\sim236$ GHz nuclear continuum spectral slopes are typically well described by standard ADAF models of low-luminosity AGN. When we subtract the unresolved emission, we detect significant extended emission in five of the seven galaxies. The extended continuum emission in the CO-bright galaxies coincides with the optical dust disks, and we determine that their extended continuum spectral indices are consistent with a thermal dust origin. However, the extended continuum features in the two CO-faint galaxies are oriented nearly perpendicular to their optical dust disks and may trace marginally resolved synchrotron jets on $<100$ pc scales.

\citet{bar16b} have carried out dynamical model fits to the \mbox{NGC 1332} ALMA \mbox{Cycle 2} CO(2$-$1) observations, and in a future paper we will model the data cubes of the four remaining CO-bright disks to derive initial estimates of (or upper limits to) their BH masses. Additional observations of molecular gas in ETGs have been obtained in \mbox{Cycle 3} (by ourselves and other groups), and this growing sample will provide additional candidates for follow-up, high-resolution study with ALMA. In a separate paper, we will further explore molecular gas and continuum properties to investigate dust disk masses and temperatures, as well as star formation thresholds, for a larger sample of ETG gas disks.

\acknowledgments

We thank the anonymous referee for helpful comments and suggestions. This paper makes use of the following ALMA data: ADS/JAO.ALMA\#2013.1.00229.S and 2013.1.00828.S. ALMA is a partnership of ESO (representing its member states), NSF (USA) and NINS (Japan), together with NRC (Canada), NSC and ASIAA (Taiwan), and KASI (Republic of Korea), in cooperation with the Republic of Chile. The Joint ALMA Observatory is operated by ESO, AUI/NRAO and NAOJ. This paper uses data taken with the NASA/ESA Hubble Space Telescope, obtained from the Data Archive at the Space Telescope Science Institute under GO programs 5910, 5999, 6094, 9427, \& 10911. The Hubble Space Telescope is a collaboration between the Space Telescope Science Institute (STScI/NASA), the Space Telescope European Coordinating Facility (ST-ECF/ESA) and the Canadian Astronomy Data Centre (CADC/NRC/CSA). Use is also made of the NASA/IPAC Extragalactic Database (NED), which is operated by the Jet Propulsion Labratory, California Institute of Technology, under contract with the NASA.

Support for this work was provided by the NSF through award GSSP SOSPA3-013 SOS funding from the NRAO and through NSF grant AST-1614212.  LCH acknowledges support from the National Natural Science Foundation of China (grant No. 11473002) and the Ministry of Science and Technology of China (grant No. 2016YFA0400702).

\bibliographystyle{apj}
\bibliography{ms.bib}

\end{document}